\documentclass[a4paper,12pt,english]{article}
\usepackage[english]{babel}
\usepackage{graphicx}
\usepackage{lnfprep}
\usepackage[top=2.5cm, bottom=3cm, left=3cm, right=3cm]{geometry}

\def\simge{\mathrel{%
      \rlap{\raise 0.511ex \hbox{$>$}}{\lower 0.511ex \hbox{$\sim$}}}}
\def\simle{\mathrel{
      \rlap{\raise 0.511ex \hbox{$<$}}{\lower 0.511ex \hbox{$\sim$}}}}
\newcommand{\Header}{
  \resizebox{15cm}{!}{
     \begin{tabular}{l r}
       \includegraphics[width=5cm, trim={0 100 0 0}]{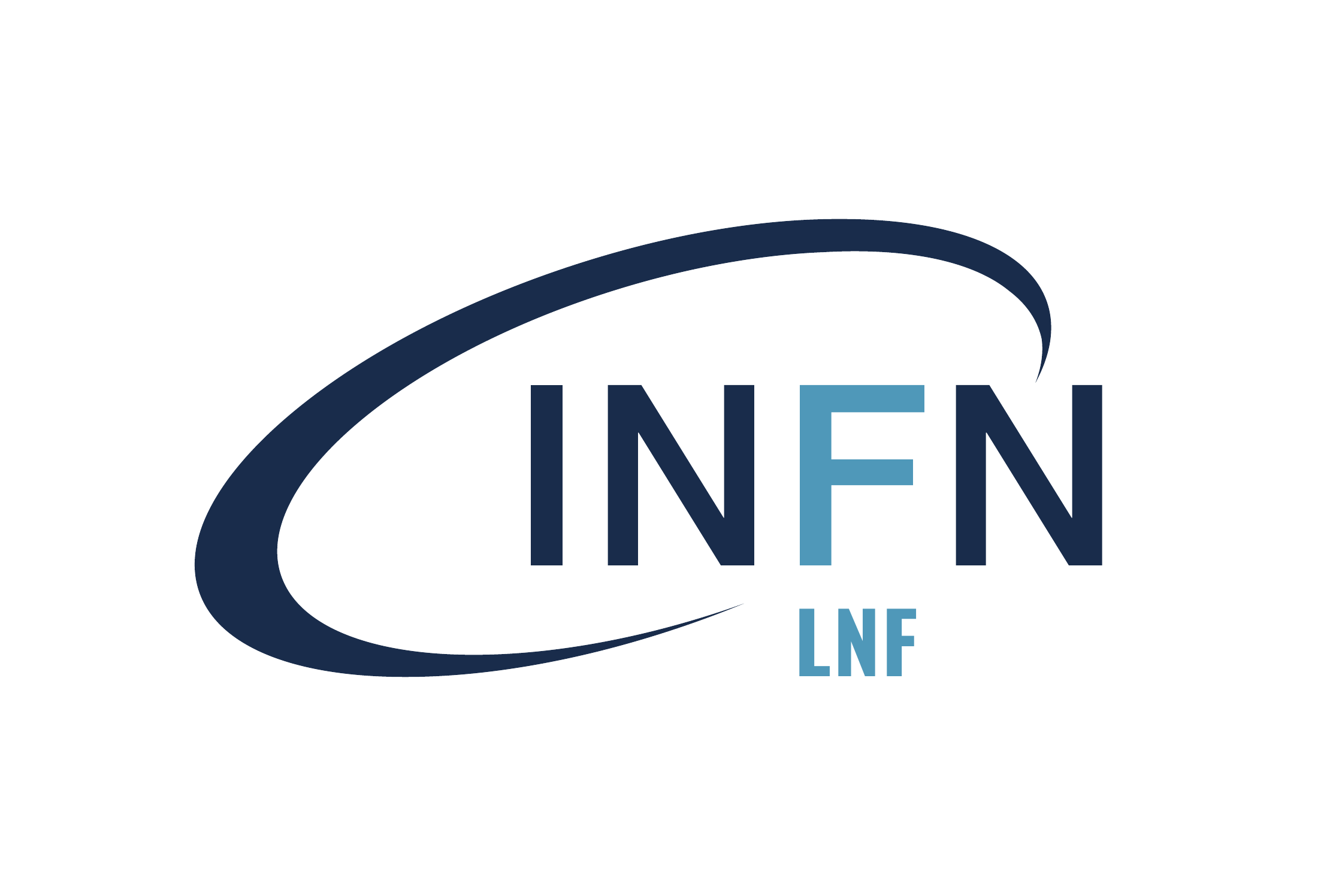} & {\LARGE\sffamily LABORATORI~NAZIONALI~DI~FRASCATI}\\
         \\
         \\
         &{\underline{\bf INFN--19-17/LNF}}\\    
         &{\small October 10, 2019} \\                      
         &MIT-CTP/5150\\
    
    \end{tabular}
    }
    \vskip 1cm
} 

\usepackage[normalem]{ulem} 
\usepackage {ulem} 

\usepackage{wrapfig}
 \usepackage{hyperref}
\usepackage{cleveref}
\usepackage{natbib}
\bibpunct[;]{(}{)}{,}{a}{}{;}
 \usepackage{type1cm}   
 \usepackage{array} 
\hypersetup{ colorlinks=true,  linkcolor=blue,  filecolor=magenta,       urlcolor=blue, citecolor=black}

\usepackage{mathptmx}
\usepackage[intlimits]{amsmath}
\usepackage{helvet}
\usepackage{courier}
\usepackage{type1cm}  
\usepackage[font={small}]{caption}
\usepackage{makeidx}         
\usepackage{multicol}        
\def\BT{Bruno Touschek}
\def\RW{Rolf Wider\o e}
\def\T{Touschek}

\def\LAL{Laboratoire de l'Acc\'el\'erateur Lin\'eaire}
\def\FJ{Fr\'ed\'eric Joliot}

\def\CdF{Coll\`ege de France}

\def\WW2{World War II}

\def\EYT{Elspeth Yonge Touschek}
\def\RLM{Reichsluftfahrtministerium}
\def\Gott{G\"ottingen}
\def\tr{\textcolor{red}}

\def\bef{\begin{figure}}
\def\enf{\end{figure}}
\def\befoot{\begin{footnotesize}}
\def\enfoot{\end{footnotesize}}
\setcitestyle{notesep={, }}

\begin{document}
\begin{titlepage}
\title{
\Header 
 {\Large \bf Bruno Touschek in Germany after the War: 1945-46 }
}


 \author{ Luisa Bonolis$^1$,
Giulia Pancheri$^2$$^,$$^\dagger$\\
{\it ${}^{1)}$Max Planck Institute for the History of Science, Boltzmannstra\ss e 22, 14195 Berlin, Germany}\\
{\it ${}^{2)}$INFN, Laboratori Nazionali di Frascati, P.O. Box 13,
I-00044 Frascati, Italy}
}  
\date{today}
\maketitle
\baselineskip=14pt
\begin{abstract}
Bruno Touschek  was an  Austrian born theoretical physicist, who proposed and built the first electron-positron collider in 1960 in the Frascati National Laboratories in Italy. In this note we reconstruct a crucial period of Bruno Touschek's life 
so far scarcely explored, which runs from Summer 1945  to the end of 1946.  
We shall 
 describe his  university studies in \Gott,   placing them   in the context  of the reconstruction of German science after 1945.
 The influence of Werner Heisenberg and other prominent German physicists will be highlighted. In parallel, we shall show how  the  decisions of the Allied powers,  towards restructuring science and technology in the UK after the war effort,
   determined    Touschek's move to  the University of Glasgow in 1947.
\end{abstract}

\vskip 2.0cm
\begin{center}
 {\it Make it a story of distances and starlight}\\
 \vskip .5 cm
Robert Penn Warren,  1905-1989,
 \copyright 1985 Robert Penn Warren
\end{center}
\vskip 2cm
\rule{15.0cm}{0.09mm}\\
\begin{footnotesize}
 {\it  e-mail:} lbonolis@mpiwg-berlin.mpg.de, pancheri@lnf.infn.it.  Authors' ordering in this and related works alternates to reflect that this work is part of a joint collaboration project with no principal author.\\
 ${}^{\dagger)}$ Also at Center for Theoretical Physics, Massachusetts Institute of Technology, USA.\\

\end{footnotesize}
\end{titlepage}
\pagestyle{plain}

\newpage
\tableofcontents
\section*{Premise}
The present note forms part of   a project that aims  to  tell the story  
of  Bruno Touschek, the maker of 
the first  electron-positron collider, named AdA, Anello di Accumulazione\footnote{Storage ring in English}, a type of elementary particle accelerator  that opened the way to present-day machines, 
\begin{wrapfigure}{r}{0.37\textwidth}
\centering
    \includegraphics[width=0.37\textwidth]{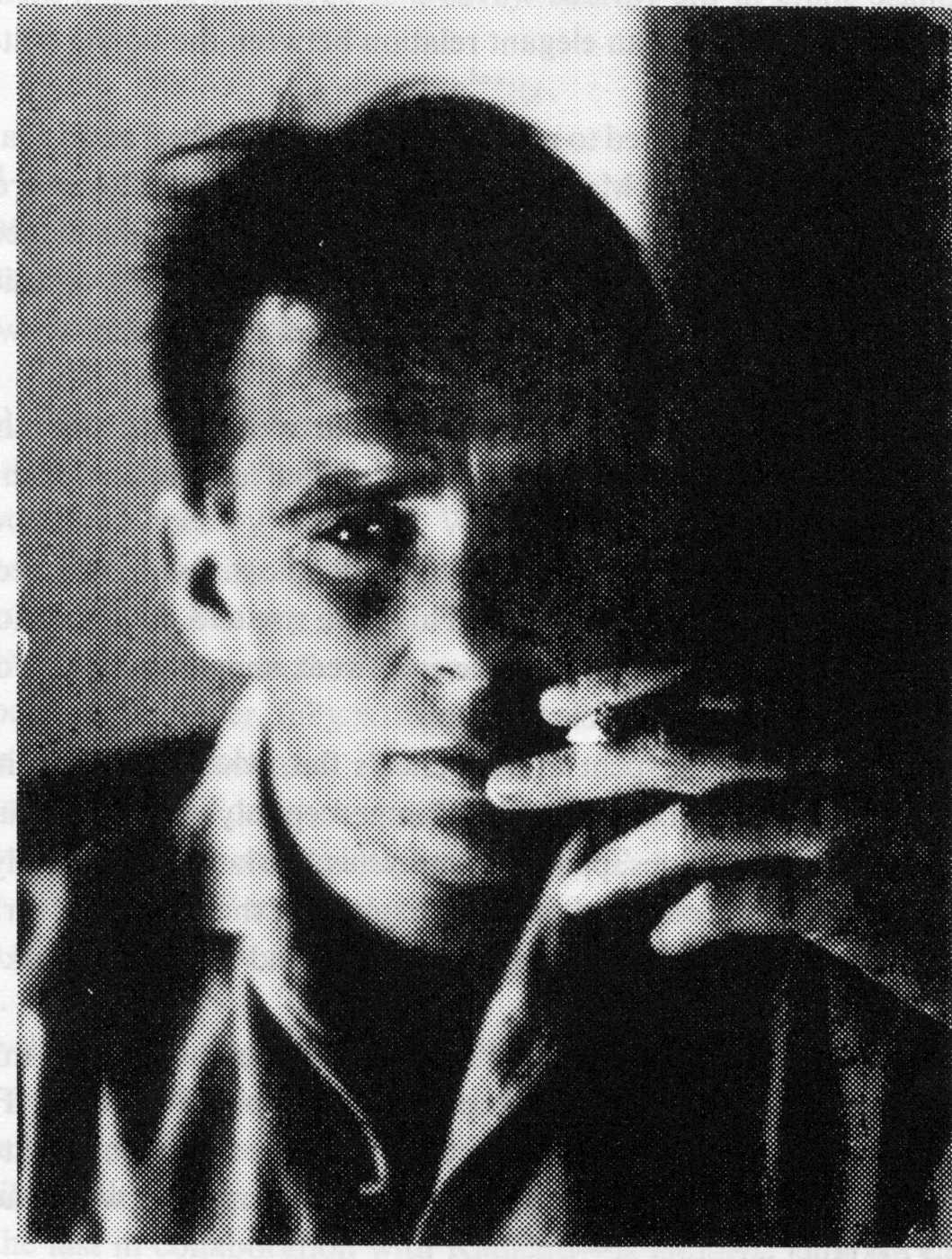}
    \caption{Bruno Touschek in 1955 from \citep{Amaldi:1981}.}
    \label{fig:Touschek-sigaretta}
    \end{wrapfigure}
such as the CERN Large Hadron Collider.

Bruno Touschek, shown here at right in a 1955 photograph, Fig.~\ref{fig:Touschek-sigaretta}, was born in Vienna, on February 7th 1921. In  view of the approaching   hundredth anniversary of his birth, 
we have   already posted two notes  which are part of this project   \citep{Bonolis:2018gpn,Pancheri:2018xdl}.
These notes cover  the years 1961-1964, during which  experimentation with AdA,  built  in Italy at  the Frascati National Laboratories,  took place, first in Frascati and  then at \LAL \ d'Orsay, in France. In what follows we go back to an earlier period of Touschek's life, the one immediately following the end of \WW2. This was a crucial transition time -- for all of Europe. During these months,  Touschek moved away from a war-related  scientific activity, spent in  semi-hiding in Germany  because of  his Jewish origin, and went on to become a physicist, under the guidance of  the great German scientists assembled in \Gott \ by the Western Allied Forces  to support  the reconstruction of European science. In a parallel move, the British scientists, who had actively  participated in the war effort  and had now engaged in  the conversion of UK physics from a war context to  new research and civil society applications,   were instrumental in favouring  Touschek's  move to the University of Glasgow in 1947. Touschek's    
  formation as   a theoretical physicist  was completed in Glasgow,  later enabling   his deep understanding of the symmetries of nature, which would lead to the   AdA project in 1960. This  period in Touschek's life will be narrated in another forthcoming note.

\section{Introduction}\label{sec:germany}
On May 7th, 1945, with the official surrender of Germany, \WW2 came to its end in Europe.\footnote{The official date for the end of the war is different from country to country. In Italy for instance, the {\it Giorno della Liberazione}, the day of  Liberation of Italy, is celebrated on April 25, which is the day the freedom fighters,  {\it i partigiani} in Italian, entered Milan,
whereas in Paris {\it la Lib\'eration de Paris}   falls on 25 August 1944,  which is the day the German command in France surrendered.  In Asia, the war ended   only after the second atomic bomb was dropped on Nagasaki, with  Japanese forces surrendering  on August 15th 1945.} The immense bloodshed and destruction that  had overcome Europe 
 were over. 

Amid the million Europeans starting on a new road to peace and collaboration,   there are the early  protagonists of  the   story of electron-positron colliders,
 the Austrian  \BT \ and   the  Norwegian   \RW. Bruno and Rolf had come together in 1943, during the  darkest times  of \WW2 and worked for two years on  the  15-MeV German   betatron, commissioned to Wider\o e by the \RLM \ (Reich Air Ministry) for, allegedly, war purposes \citep{Amaldi:1981,Wideroe:1994,Bonolis:2011wa}. 
   
  Before, during and after the war, many pathways criss-crossed Europe to ultimately lead to  the construction of AdA, the first ever electron-positron collider,   built in Italy  in  1960.  Brought to Orsay in 1961,  a Franco-Italian team under Touschek's guidance proved its feasibility as a major tool to explore the world of elementary particles.
  When Bruno and Rolf met,   two of these pathways  came together.  One road came from Norway, the other from {Central Europe},
  Austria and Germany.
Then, after the war, the destinies of the two scientists took different ways.   
Between March and April 1945, as described in \citep{Wideroe:1984aa}, \citep{Brustad:2009aa}, and more in depth in \citep{Sorheim:2015},  \RW \ returned  to Norway, where, in May, shortly after the German surrender, he was arrested  and accused of having worked on the development of V2 rockets.\footnote{See letter from Wider\o e to Ernst Sommerfeld from Baden, dated April 12, 1946. Deutsches Museum Archive, NL 089, box 014: ``My question whirled up naturally a lot of dust and quite fantastic things were hypotesized  (for example, I was supposed to have invented  the V2)'' (Meine Sache wirbelte nat\"urlich sehr viel Staub auf und man vermutete ganz phantastische Sachen (beispielweise sollte ich angeblich die V2 erfunden haben)).}
He wrote an extensive report on his work on the betatron construction in Hamburg and  was released in July, but only in February 1946 it was clarified that his work had not been of any military value to Nazi Germany. However, he was burdened with financial penalties and was eventually allowed to move to Switzerland, where he took up a leading position at Brown Boveri \& Co and  applied  his knowledge of accelerator science  to medical developments.
   
As for Touschek, 
 his mind and heart were now bent on  regaining the lost years and finishing the studies he had started at the University of Rome in Spring 1939, after passing his high school exam, the {\it matura}, at the  Staatsgymnasium I in Vienna in February 1939 \citep{Amaldi:1981}.
In those years, after the   {\it Anschluss}, {namely the  annexation of Austria to Germany,  and the promulgation of } the N\"uremberg laws,   his Jewish origin on the maternal side had  derailed  his life and  studies.\footnote{On March 12th 1938, Austria was annexed to Germany, and on May 25th the German N\"uremberg laws, affecting citizens of Jewish origin, were applied to Austrian citizens as well. These laws distinguished between various degrees of Jewish parentage. For a poignant memory of Vienna during   the Anschluss, see \citep{DeWaal:2012}.}
 After an attempt to emigrate to England and study chemistry at the  University of Manchester,\footnote{Letter to parents (father and stepmother),  20 March 1939, from Rome.} he had enrolled in physics at the University of Vienna   in fall 1939, but, at the end of the academic year, he had been expelled from the University because of his Jewish origin.     To continue his studies, as other possibilities were now closed to him, he moved out of Vienna, to Germany, where  his  (non-Jewish) last name could allow him to move around, {\it incognito}. There,  in  Hamburg and Berlin, from 1942 until 1945,  he attended physics lectures, leading the life of a student without being one,  trying not to be noticed by the omnipresent Gestapo. 
 Once the war was over, at the end of 1945,  he finally could  try to fulfill his dream to become a physicist.
After one   year in G\"ottingen, Touschek went to Glasgow for 5 years, for his doctorate and three years of lecturing and research, and then moved to Rome, hired by Edoardo Amaldi, one of CERN's founding fathers, to do research in cosmic rays, theoretical physics and assist experimentalists in accelerator activities.  When he   arrived in Italy, in 1952, he  joined  the road of postwar reconstruction of Italian physics. This road, and the synchrotron, which would be built on the gentle slopes of the ancient hills overlooking Rome from the South-east, in the newly founded National Laboratories  in Frascati, would bring him to propose the construction of AdA, the first ever accelerator for  electron-positron collisions,  an experiment, he said,  ``really worth doing".\footnote{See Minutes of Frascati Laboratories Meeting of Febuary 17, 1960 in \citep{Pancheri:2018xdl}.}

In what follows, we shall tell the story of Touschek in the transition period in G\"ottingen, 
 from the end of the war in 1945 through 1946, when he was getting ready to join  the University of Glasgow as a doctoral student. 
The present note  
 focuses on a period of Touschek's life, so far not much explored.  Starting with Bruno in Wrist in 1945 soon after the war ended, we shall go  beyond what is known from   \citep{Amaldi:1981} and highlight the  relevance of this early  post-war  period in Touschek's formation as physicist, through  the impact of  the German  experience, first as a Diploma student  and then as a researcher  in G\"ottingen. 

We shall start by recalling Touschek's life during the last months of the war, relying  on the two letters, sent by Bruno Touschek to his parents in 1945, already published in \citep{Bonolis:2011wa},  and which are part of a copious correspondence Touschek  kept up  with his family throughout his life. For the period to follow, our reconstruction  
is mainly based 
on documents from a number of public archives, such as, in particular,  Edoardo Amaldi and Bruno Touschek Archives in Rome University,  Arnold and Ernst  Sommerfeld Archives at the  Deutsches Museum in Munich, and Werner Heisenberg's papers at the Archives of the Max Planck Society in Berlin. In addition, we have  consulted  family documents to which we were given access by Bruno Touschek's widow,   the late  \EYT, during a series of encounters between 2003 and 2011.  


We had  both been acquainted with Elspeth    for quite  some time. In 1966 and 1967, one of us, G.P., had been  a researcher  with Touschek's group at Frascati National Laboratories and   had occasion to meet   her,  first in Rome, and  later during  a  vacation in Positano. It was September 1966, and Bruno had invited the  young researchers from  his theory group, Paolo Di Vecchia, Giancarlo Rossi and G.P., to join him and the family  for a few days of sun and swimming on the Amalfi coast \citep{Greco:2005}.  

In May 1978, Touschek passed away. A few years later,  G.P. had met again Elspeth and her son Francis  in 1987, during the first edition of the \BT\ Memorial Lectures, held in the INFN National Laboratories in Frascati. Later on, the other author of this note, L.B., interested in preparing a docu-film about  Touschek on the occasion of 25 years since his death, started visiting Elspeth, who had now left the city  to live in   the countryside. Elspeth showed to L.B.   letters, photographs  and other documents, ranging from {\it circa} 
1936 
to 1971, and which were used in part for the  docu-film \href{https://youtu.be/R2YOjnUGaNY}{\it Bruno Touschek and the art of physics}, authored by E. Agapito and L. Bonolis in 2003, and in \citep{Bonolis:2005}.  

In  2008, the  two of us started  preparing an article about Touschek's life and accomplishments \citep{Bonolis:2011wa},  which could  update  Amaldi's biographical work \citep{Amaldi:1981}. We  went  to visit  Elspeth together, and she shared with us some of her memories as well as  additional letters and  documents  about Touschek's life, which were on her possession and which she had carefully organized in chronological ordering. 
The present narration, if not otherwise indicated, is based on our conversations with her and the material to which she  gave us access. Before starting our narration, we wish to express our deep gratitude to Elspeth  for bestowing on us her friendship. 
  

\section{ Hamburg 1945: from death rays to post-war science }
\label{sec:1945}
Wider\o e's  betatron had been built in Hamburg,  at the C.H.F.  M\"uller factory. But as the war entered into its final months, it became clear that it could  be destroyed by the  Allied heavy bombing or captured by advancing enemy troops, either the Western Alliance or the Soviet Army. 

\begin{wrapfigure}{l}{0.37\textwidth}
\centering
    \includegraphics[width=0.37\textwidth]{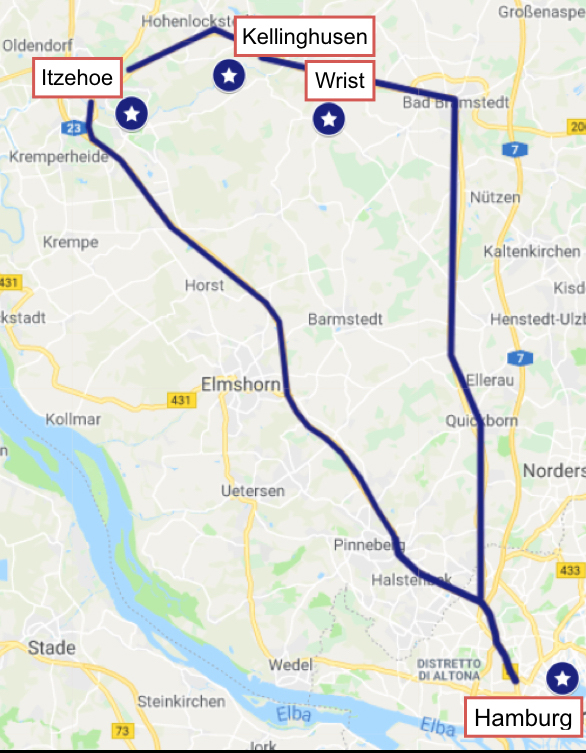}
    \caption{Touschek's movements in March 1945.} 
    \label{fig:Kellinghusen}
\end{wrapfigure} 
In March, the German Aviation Ministry ordered the betatron group to  bring it to a safer location. A disused factory in Kellinghusen, in the surroundings of Wrist, a few kilometers  north of Hamburg,   was found, and the transport was over by mid-March.
In Fig.~\ref{fig:Kellinghusen} we show a map locating Touschek's movements in March 1945, as described in the October and November  letters to his parents \citep{Bonolis:2011wa}.

Immediately after the betatron was away and safe,  Touschek was arrested by the Gestapo.  This is what he had feared already two years before, when, 
 in fall 1943, he  had joined Wider\o e's   classified betatron project.   \T\   
saw  that such involvement would expose him to the Gestapo, who would make inquiries and thus  learn of his Jewish origin.\footnote{Letter to  parents, 29th October 1943, from Berlin.}
What he had feared   was now happening: his usefulness over, the Gestapo  was ready to arrest him, eventually sending him later  to a concentration camp,  as had happened to many technical or scientific employees in similar  condition.\footnote{Amaldi, probably quoting Touschek, mentions his reading foreign papers in a public library, as the reason for his imprisonment  \citep{Amaldi:1981}. A slightly different story comes from Carlo Bernardini, Touschek's closest friend in Rome. According to him,  Touschek was seen drawing gyroscopes in some public place, and since gyroscopes could be related to rockets (V-2) control system, he was  accused of espionage. Both stories could come from Touschek himself, either reading of foreign magazines and drawing of gyroscopes could be true, but the explanations seem contrived. Touschek's letters of the Fall 1943  suggest a  much darker and dramatic explanation: when Touschek signed his contract to work on the betatron project, under the \RLM's  control, his life was obviously investigated and his Jewish origin became known.} The events  of his imprisonment and those immediately following it, have been part of Touschek's legend. Their general outline was presented by Edoardo Amaldi, who heard the story from Touschek himself, and included it in \citep{Amaldi:1981}. The details are now known from two letters he wrote to his parents, the first on June 22nd and the second on November 17th  1945, published in English translation in  \citep{Bonolis:2011wa}. 
From these, one learns, first hand and just as  memories were still vivid and precise, the sequence of almost miraculous events  which allowed Touschek to escape death and which we summarize below for convenience  of the reader.\footnote{A full description of Touschek's   whereabouts  from mid March to November 1945, are to be found in the   two letters  first  published in their entirety in \citep{Bonolis:2011wa}. Quoted remarks in this section refer to  these  letters.}

In the October and November  letters, \T\ recounts how 
he was held in the infamous Fuhlsb\"uttel prison near Hamburg from March 15th until mid April 1945. After a first week of hardship and despair, during which he even considered suicide, he was able to receive visits from \RW, who was still in Germany.\footnote{For \RW's movements after he left Germany sometimes in April 1945, see \citep{Wideroe:1994}.}
Bruno  was   reassured he could soon be free,  as it was being  clarified that his work was very important for the betatron project. On the contrary   April came, and he was still held prisoner. Then, as the Allied forces were approaching Hamburg,  orders came for the 200  prisoners from  Fuhslb\"uttel to be moved out  towards the Kiel concentration camp, $\sim$ 100 km North. Touschek was one of them. 

On the way to Kiel,   Touschek, who was ill and carried a heavy package of books, fell to the ground. As he wrote to his parents \citep{Bonolis:2011}:
\begin{quote}
\dots I definitely broke down in Langenhorn. 
\end{quote}

\begin{figure}[htb]
\centering
\includegraphics[scale=0.31]{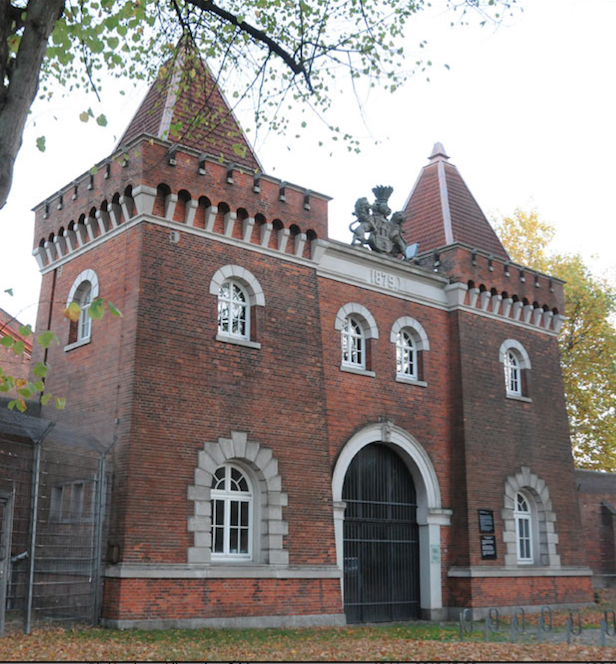}
 \includegraphics[scale=0.4]{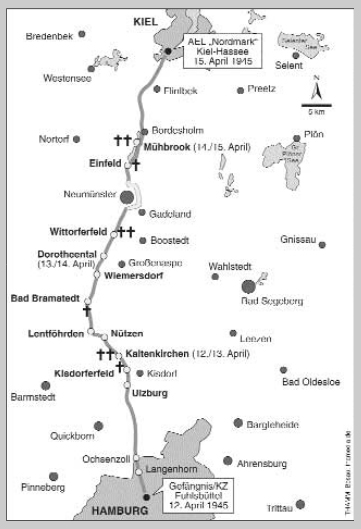}
\caption{At left,  a contemporary photo of the entrance to the Fuhlsb\"uttel prison in Hamburg where Touschek was held for about 6 weeks in Spring 1945. At right, a map of the forced march from Fuhlsb\"uttel to Kiel, which brought  the 200 prisoners from Hamburg to Kiel, between 12-14 April 1945, from \citep{Fentsahm:2004aa}.  Langenhorn, where Touschek was shot and fell to the ground, while the column continued towards Kiel without him,   is seen at bottom, within the Hamburg region.  The crosses indicate prisoners' deaths.}
\label{fig:mapHamburg}
\end{figure}

He was then shot by one
 of the SS guards escorting the prisoners, and left for dead. The forced march from Fuhlsb\"uttel to Kiel is described in \citep{Fentsahm:2004aa}, where one can also find the map shown in Fig.~\ref{fig:mapHamburg}.
 
 As he regained consciousness, he was first brought to a hospital and then through ``all kind of prisons", the last being in  Altona, 
 in the surroundings of Hamburg \citep[5,7]{Amaldi:1981} \citep[45]{Bonolis:2011wa}.

Those were the final days of the war in Europe, indeed the last hours, during which    prisoners risked  being  killed, often to prevent  witnesses from  surviving. Touschek was lucky, or perhaps, and  more likely, the tight grip held by the Nazis was  at its  end. On April 30th, Theodore Hollnack, the administrator of Wider\o e's betatron project, eventually came to  free Touschek.\footnote{In a letter to his friend Ernst Sommerfeld describing all this, Touschek wrote that he  was angry because Hollnack had waited too much: ``He then explained to me -- after having done nothing for three weeks -- that without him I would have definitely been shot." Deutsches Museum Archive, NL 089, box 014.}
 
Two days earlier, the British army had started the final assault on the city of Hamburg, where the German command was holding  against the Allied invasion, and the fight moved from block to block through the city.  The city surrendered on May 3rd.\footnote{An eerily silent footage about the entrance of the Allied troops in Hamburg can be found in \url{https://www.youtube.com/watch?v=en3hkuc1QoM}. See also details about the  \href{https://en.wikipedia.org/wiki/Battle_of_Hamburg_(1945)}{ Battle of Hamburg-1945}.}
 
 \begin{figure}
 \centering
 \includegraphics[scale=0.225]{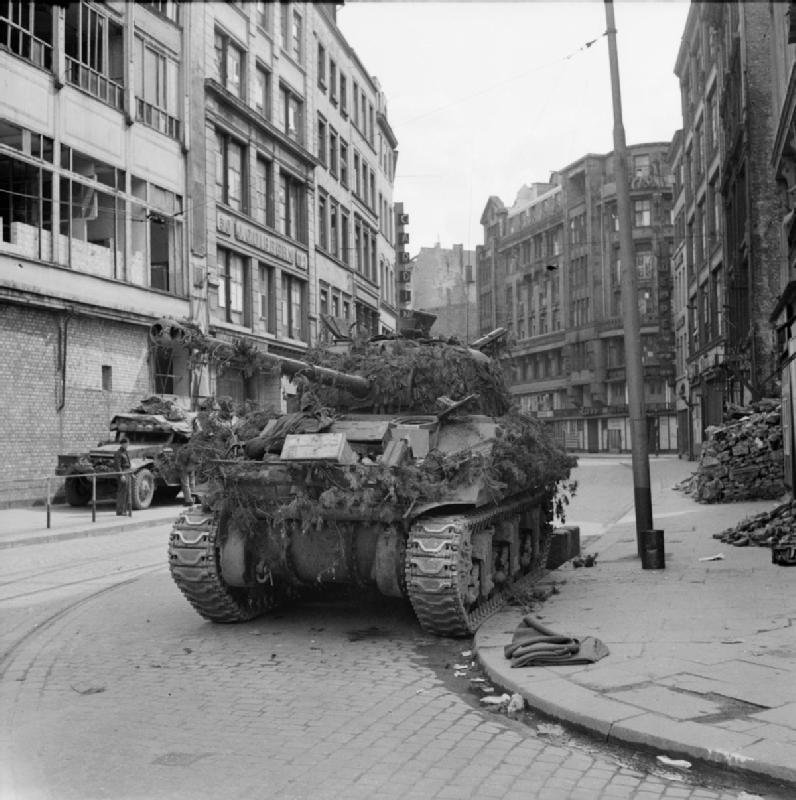}
 \includegraphics[scale=1.2]{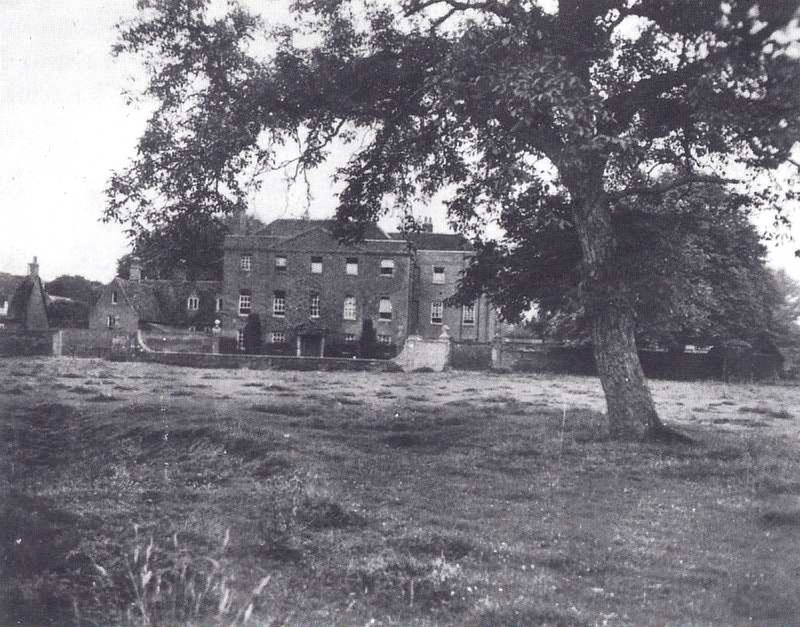}
 \caption{At left, an  image of the battle for Hamburg, May 1945,  from  \url{https://en.wikipedia.org/wiki/Battle_of_Hamburg_(1945)}.
 At right,  Farm Hall in Godmanchester, England, where the Uranium scientists were held {\it incommunicado} for 7 months, from \url{https://en.wikipedia.org/wiki/Operation_Epsilon\#/media/File:FarmHallLarge.jpg}.}
 \label{fig:hamburg1945}
 \end{figure}
\section{German science and the mission of the T-force}
\label{sec:T-force}
The last year of the war saw not only the heavy   bombing of German cities and installations, but also   planning for the future of the Western world, as it came to be called. The  position of eminence of Germany in science and all fields of technology in Europe had been such that, as the various Allied armies progressed through Germany, they  raced
    to  secure what would be the  most prized  booty, depending on the stage of scientific and technological advancement of the different countries.
    What the Germans had achieved  in science and technology  since the late 1930's, would be important to know and to acquire in view of  the new world political  assessment after the war. To this end, 
the Western Alliance set in motion  a number  of different operations which would lead to the capture of a vast amount of German industrial, scientific  and technical equipment   as well as of the most prominent German scientists, who were quickly transported to the United States and to England.\footnote{For a good journalistic overview, see  \url{https://www.theguardian.com/science/2007/aug/29/sciencenews.secondworldwar}. } 

All along, even  before the final surrender of Germany, a special task force under joint American-British command, named the T-Force,  had been scouting Germany for its industrial and scientific resources, racing to reach Germany's top scientists before the arrival of the Russian Army 
 \citep{Bernstein:2001} \citep{Longden:2009aa}. One of the key actions of T-Force units was the Allied Scientific Intelligence Mission, code-named ``Alsos'', brain-child of Colonel Leslie Groves, the military head of the Manhattan Project.\footnote{``Alsos"  is also the Greek word for `grove'.} The Alsos Mission, headed by U.S. Army Lieutenant Colonel Boris T. Pash, was set up to seize key elements of the German nuclear energy project working at Hechingen, in southwest Germany, where the Kaiser Wilhelm 
Institute for Physics had been evacuated from Berlin-Dahlem  \citep{pash:1969} \citep{goudsmit:1996} \citep{Cassidy:2017}.\footnote{Efforts in Hechingen had been concentrated on trying to achieve criticality in a primitive research reactor they had assembled within a cave in the nearby town of Haigerloch.} Other actions concerned the rocket scientists, as well as biological and chemical warfare experts \citep{Jacobsen:2014}.\footnote{For the Americans  a major  target became the rocket scientists, foremost among them   Wernher von Braun, who had been in charge of the nazi  V-2 program in Peenemunde, and would  later bring   the US to  land on the Moon.} And, as  we shall see in \ref{ssec:tforcebetatron},  particle accelerators as well would be  a target of interest.  
 \subsection{Operation Epsilon}
 Operation Epsilon was the codename of the program in which the main protagonists of the German nuclear program -- the Uranium Club -- were flown to England at the beginning of July and held in secrecy at the country estate Farm Hall, near Cambridge. A primary aim of the program was to understand how close Nazi Germany had been to building a nuclear bomb by listening to their conversations through hidden microphones \citep{Bernstein:2001,Cassidy:2017,mcpartland_farm_2013}.\footnote{The 10 leading German nuclear scientists brought to Farm Hall were Erich Bagge, Kurt Diebner, Horst Korsching,  Walter Gerlach, Otto Hahn, Paul Harteck, Werner Heisenberg, Max von Laue,   Carl F. von Weizs\"acker, and Karl Wirtz. The American officers in command wanted to keep them under constant guard, as prisoners, but the British captain in charge explained there would be no need for this, if the scientists could be convinced to give their word of honour not to escape.
The transcript of these conversation were held classified until 1992, when they were released following a request   addressed to the House of Lords by the  President of the Royal Society in London. In the letter, to whose draft contributed Rudolf Peierls, later external PhD supervisor of Touschek in Glasgow,    the following 1985 words by the German President Richard von Weizsa\"cker,  are quoted: ``We need and we have the strength to look truth straight in the eye without embellishment and without distortions [\dots] anyone who closes his eyes to the past is blind to the present".  The British version of the transcripts  is  now available for free download  at \url{http://discovery.nationalarchives.gov.uk/SearchUI/Details?uri=C4414534}. For the  US copy, see   NARA, RG 77.11.1 (Office of the Commanding General), entry 22, Box 163, in   \citep{Cassidy:2017}.  An Italian translation was published in 1994 \citep{epsilon:1994}.}

 Among  the Farm Hall detainees, there  was   Werner Heisenberg, one of the founders of Quantum Mechanics and one of the most illustrious German scientists,  a key theoretical figure in the German nuclear project,  and, since 1942, official director of the Kaiser Wilhelm Institute for Physics in Berlin-Dahlem {\citep{Cassidy:1993aa}}.\footnote{For an  extensive bibliography of Heisenberg's work, see \url{http://www.netlib.org/bibnet/authors/h/heisenberg-werner.html}. For the collected works see as well \citep{Heisenberg:1983aa}.} He was taken into custody by Alsos on May 4th, in the little village of Urfeld, where he had arrived after a desperate bicycle ride across war-torn southern Germany to reach his family, as soon as news of the French advance had reached Hechingen.

 \begin{figure}
 \centering
 \includegraphics[scale=0.43]{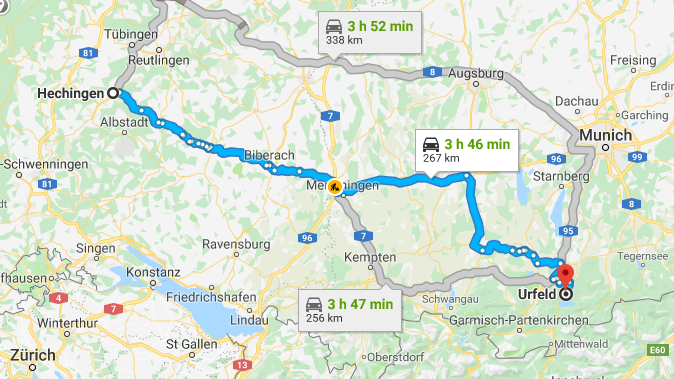}
 \includegraphics[scale=0.07]{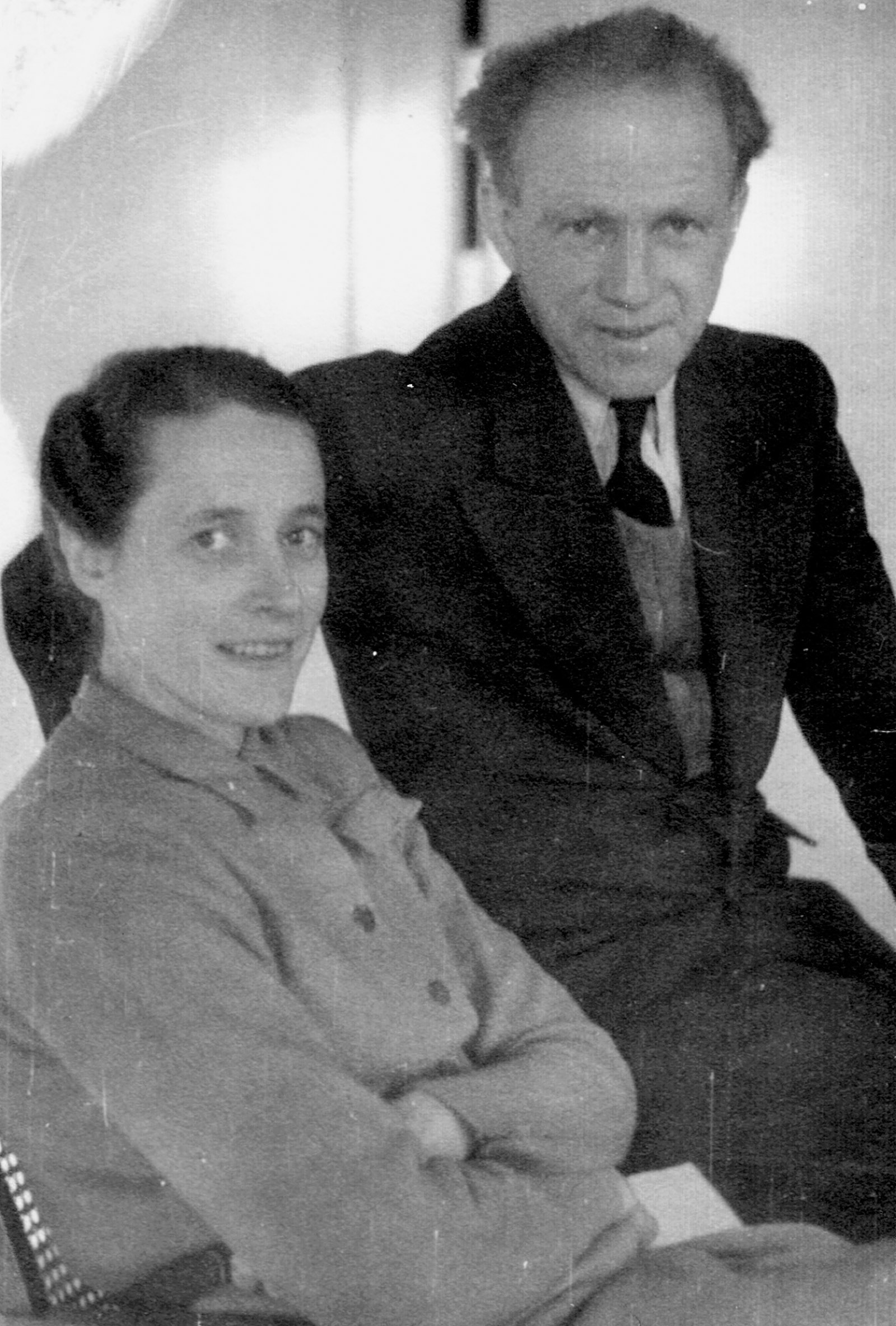}
 \caption{At left, a contemporary map  showing the long trip Heisenberg took to reach his family, as the Allied forces were progressing through Germany,  from Hechingen to Urfeld, where he was taken charge by the ALSOS mission on May 4th 1945. At right, a (circa) 1946 photograph of Werner and Elizabeth Heisenberg, Heisenberg Family Archives.}
 \label{fig:germanyheisenbergflight}
 \end{figure}

 In Heisenberg's own words \citep[190-192]{Heisenberg:1971aa}:
   
\begin{quote}
In the middle of April [1945], the last German stragglers passed through Hechingen heading East. One afternoon we could hear the first enemy tanks. In the South, the French had probably advanced well past Hechingen, as far as the ridge of the Rauhe Alb. It was high time I was gone. Toward midnight,   Carl Frederick [von Weizs\"acker] returned from a bicycle reconnaissance tour of Reutlingen. We held a small farewell celebration in the air raid shelter of the institute and at about 3 a.m. I set off in the direction of Urfeld [\dots] 

[\dots] it was not until three days after I had set out that I reached Urfeld and found my family  well and unharmed  [\dots] 

On May 4th, when Colonel Pash, leading a small US detachment, came to take me prisoner, I felt like an utterly exhausted swimmer setting foot on firm land.
\end{quote}

The imprisonment of the German ``atomic scientists''\footnote{What is now called {\it nuclear physics}, and which includes studies for nuclear energy uses, both civilian and military, was then called {\it atomic physics},  hence the still used term {\it atomic bomb} and, in the context of German scientists at Farm Hall,  {\it atomic scientists}.} marked the ``zero hour'' of  the post-war future of German science.  During their seclusion in Farm Hall   which lasted 6 months, the {Western Alliance} debated and planned the reconstruction of Europe.
 In particular,  a major interest to the British political establishment was the reconstruction of Germany on its industrial and technological aspects, all of which had to start with rebuilding a strong scientific terrain.

\subsection{The T-Force and Wider\o e's betatron}
\label{ssec:tforcebetatron}
Wider\o e's betatron  was also one of the targets of the T-Force. 
Particle accelerators had now morphed from  the planning and invention stage into   the  most prominent  research tool in atomic and nuclear physics, and would become  of future strategic interest, in the mind of politicians and the military.
 \footnote{The  field of particle accelerators had  initially developed through electrostatic accelerators, most prominent of them the one by  Van de Graaff  in the United States (1929-30), followed by the   British developments by J.D. Cockcroft and E.T.S. Walton in 1932. See original papers in \url{https://royalsocietypublishing.org/doi/pdf/10.1098/rspa.1932.0107} and   \url{https://royalsocietypublishing.org/doi/pdf/10.1098/rspa.1932.0133}. New directions   had also  arisen  through the seminal work by \RW, first by proposing  the induction accelerator,  an accelerator for electrons (beta rays)  later to be called the  {\it betatron}, and then with the construction of the first linear accelerator,  the result of his doctoral dissertation in Aachen, completed in 1927 \citep{Wideroe:1928flt}. While Wider\o e had not succeeded in making his betatron work, his  linear accelerator  inspired E.O. Lawrence to build the first cyclotron, an accelerator of protons,  at the University of California in 1930, as Lawrence himself recalled in his Nobel Lecture (\url{https://www.nobelprize.org/prizes/physics/1939/lawrence/lecture/}). See also P.F. Dahl in  \citep{Dahl:1992aa} at  \url{lss.fnal.gov/archive/other/ssc/sscl-sr-1186.pdf}.  In 1933, a patent for a betatron was filed in Germany by Max Steenbeck. Later, in the 1930's, cyclotrons were also built  in Europe: among them, and of later relevance to our  story,  the one at  the \CdF\ in Paris, by \FJ \ and one in Germany by Walther Bothe and Wolfgang Gentner.} 
 As the war started,  a major advancement in the field had taken place in the United States,  with the successful  operation of a betatron, announced by Kerst in 1941 \citep{Kerst:1940zz,Kerst:1941zz,PhysRev.60.53}.\footnote{In his first article, Kerst cited \RW's seminal paper  \citep{Wideroe:1928flt}, but with a wrong year, 1938.}  This series of articles  revived the interest in betatrons and new projects were  submitted   to the German  military, which saw  them as   possible sources of deadly X-rays \citep{Waloschek:2012}. Among them,  there was  \RW's   proposal  \citep{Wideroe:1994} for  the construction of  a 15-MeV betatron -- the first at this energy in Germany -- and a parallel proposal for a much more powerful 100-MeV machine which was never built.\footnote{For a T-force report on history of betatron development, see  B.I.O.S. Report n. 77 in \url{http://www.cdvandt.org/fiat-cios-bios.htm}.}  
It is therefore quite understandable that, as the war ended, the  German knowledge  in accelerator science became of possible interest to the Allied nations, 
and in particular to the British, less so to the   Americans, whose  expertise and dominance were not lacking  in this field. The  German work on betatrons, which had been going on through the war,  became part of the British war spoils \citep{Hall:2019aa}.

After Hollnack freed Touschek from prison, they 
went back to Kellinghusen, near Wrist, where, in mid March, a time which now  seemed like centuries ago, Touschek and Rolf Wider\o e had brought the 15-MeV betatron. 
  In Kellinghusen, Hollnack  had immediately put himself at the  disposal of the British authorities, and reorganized activities around the betatron creating a small enterprise  called   the MegaVolt Forschung Laboratorium, MV-Research Association (MVRA), which gathered all the key members  of the betatron group -- previously 
  working under the guidance of  Wider\o e (the Megavolt Versuchsanstalt) -- and others.\footnote{November 17th, 1945, letter by \BT \ to his parents in \citep{Bonolis:2011wa}.} Hollnack asked Touschek to join in. Everybody was trying to survive and, for some, as in Hollnack's case,  even to strive. 
At that moment Touschek had no alternative and accepted the offer; he had been freed by Hollnack, thus avoiding 
 being killed
 in the last days of fights around the city.
 While still trying to recover from the painful memories of his losses and the trauma of imprisonment, Bruno  immediately made plans for a doctorate, as he  told his parents in his first letters written in June, where he recounted the whole story of his arrest. In the meantime, thanks to his knowledge of English,\footnote{Touschek had learned English when in Rome, in Spring 1939, when he was applying for a Visa  to go to England and  study at the University of Manchester.} 
he acted as interpreter and was then able to negotiate with the   T-Force  and have the MVRA ``occupied''   
 by them, mainly meaning being protected by the British troops against looting and other killings. 

However, he was soon out of  empathy with the group. He did not like Hollnack, nor `his grandiose ambitions', and was also eager to return to  the theoretical physics studies he had  started during his  last two years in Vienna, in  1940-41,  fostered by  his physics mentor, Paul Urban, a young assistant professor at the University of Vienna.\footnote{Paul Urban (1905-1955) assisted \BT \ to continue his studies at the University of Vienna, after June 1940, when, because of his Jewish origin,  Bruno was not allowed to follow courses anymore nor  to borrow books from the library. Later Urban   was instrumental in introducing  Touschek to Arnold Sommerfeld \citep{Amaldi:1981}. For Urban's life and scientific accomplishments, see also \citep{Guardiola:1996}.}
His dream of becoming a  physicist had then been  reinforced   by the correspondence with Arnold Sommerfeld in 1941 and the lectures by  prominent scientists he had attended during the war at Hamburg and Berlin Universities.\footnote{Sommerfeld's correspondence with Touschek referred to in this note is kept in the Archives  of the Deutsches Museum in Munich, Arnold Sommerfeld Papers, folder NL 089,013.}

By end of June \T\  asked his colleagues in the MVRA to put an end to his collaboration. They agreed that he would have a three-months leave and in late August, as he wrote in a letter to Arnold Sommerfeld dated 28 September, he went to visit some of the scientists whose lectures he had attended {\it incognito} during the war: Hans D. Jensen -- one of the members of the Uranium Club, now in Hannover  -- and Hans S\"uss in G\"ottingen, who had participated in German nuclear research activities during the war, 
 and whom he knew since his Hamburg days, as well.\footnote{At that time, both collaborated on the nuclear shell model, for which Jensen would later be awarded one half of the 1963 Nobel Prize jointly with Maria Goeppert-Mayer (the other half was awarded to Eugene Wigner for his fundamental work on symmetry  principles).} In 
G\"ottingen,  Touschek also saw  Friederich Georg Houtermans, or Fritz,  or Fissel, as he was also known, 
with whom he would remain friends until Houtermans' death in 1966.\footnote{Fritz Houtermans had  arrived in G\"ottingen in spring 1945, after a tortuous trajectory of persecution by both Nazis and Communists. Fritz Houterman's  life was the  subject of various books, in particular of an unpublished  manuscript on which Edoardo Amaldi was working before his sudden death in 1989. In 2010,  the manuscript was donated by Amaldi's family to the Bern University Laboratory for  High Energy Physics. It was then  edited  by  S. Braccini, S. Ereditato and P. Scampoli, three researchers from the University of Bern, in recognition of Houtermans' contribution to the development of particle physics in Bern, and  in Switzerland \citep{Amaldi:2012}. In the Preamble  to the unfinished book, Amaldi writes : `When in 1937 my friend George Placzek arrived in Rome from U.S.S.R, he had mentioned Houtermans as one of the young physicists gone to Karkhov to participate in the construction of a socialist society and recently in serious political troubles. I received a letter from him, from Berlin in 1942, after, as I learned late, he had succeeded in getting out of a prison to which he had been transferred from the Lubyanka in Moscow."}

Touschek's principal worry now was to formally complete  his studies, first by obtaining a degree in Physics, namely the title of Diplom-Physiker, and then continue with a doctoral thesis.  During the visit,  Jensen promised Touschek he would arrange for a PhD work and a position as assistant and he also received  a similar offer from Hamburg, where Wilhelm Lenz, director of the Institute of Theoretical Physics, who had always protected him during the war, could now openly support him to complete his university studies and continue with a PhD dissertation.

A way to proceed was in sight, and, returning to Kellinghusen,  Touschek was now eager to start writing a dissertation on the betatron.  However, {this  could not happen}. As he  would later write to his parents in the  already mentioned November 17th letter \citep{Bonolis:2011wa}:
\begin{quote}
A reunion with the T-Force has decided that things should remain a state secret so that its use for a thesis is out of the question. I will be able to leave Kellinghusen only after an Allied Commission has  decided in regard to the betatron.
\end{quote} 
Writing  to Sommerfeld in September, Touschek says that he  ``felt like a T-Force prisoner".
\footnote{Touschek to A. Sommerfeld, 28 September 1945 from Kellinghusen, Deutsches Museum Archive, Arnold Sommerfeld papers, folder NL 089,013.} Indeed, he was. In this second half of the year 1945, the Allies were making a thorough survey of the scientific achievements of German science and technology, and nothing could really start in Germany until the decisions had been taken as to Germany's future. Not unlike the members of the Uranverein (the Uranium Club), who were  held in England in Farm Hall, so was Touschek held in Germany. Unlike them, however, he  was  free to move within the British zone, still he could  not  go to Austria or  publish anything about the betatron. In the meantime he continued his work on different  theoretical topics related to the betatron, in particular on radiation damping,  but also on neutrino theory.

\begin{figure}[htb]
  \centering
\includegraphics[scale=0.31]{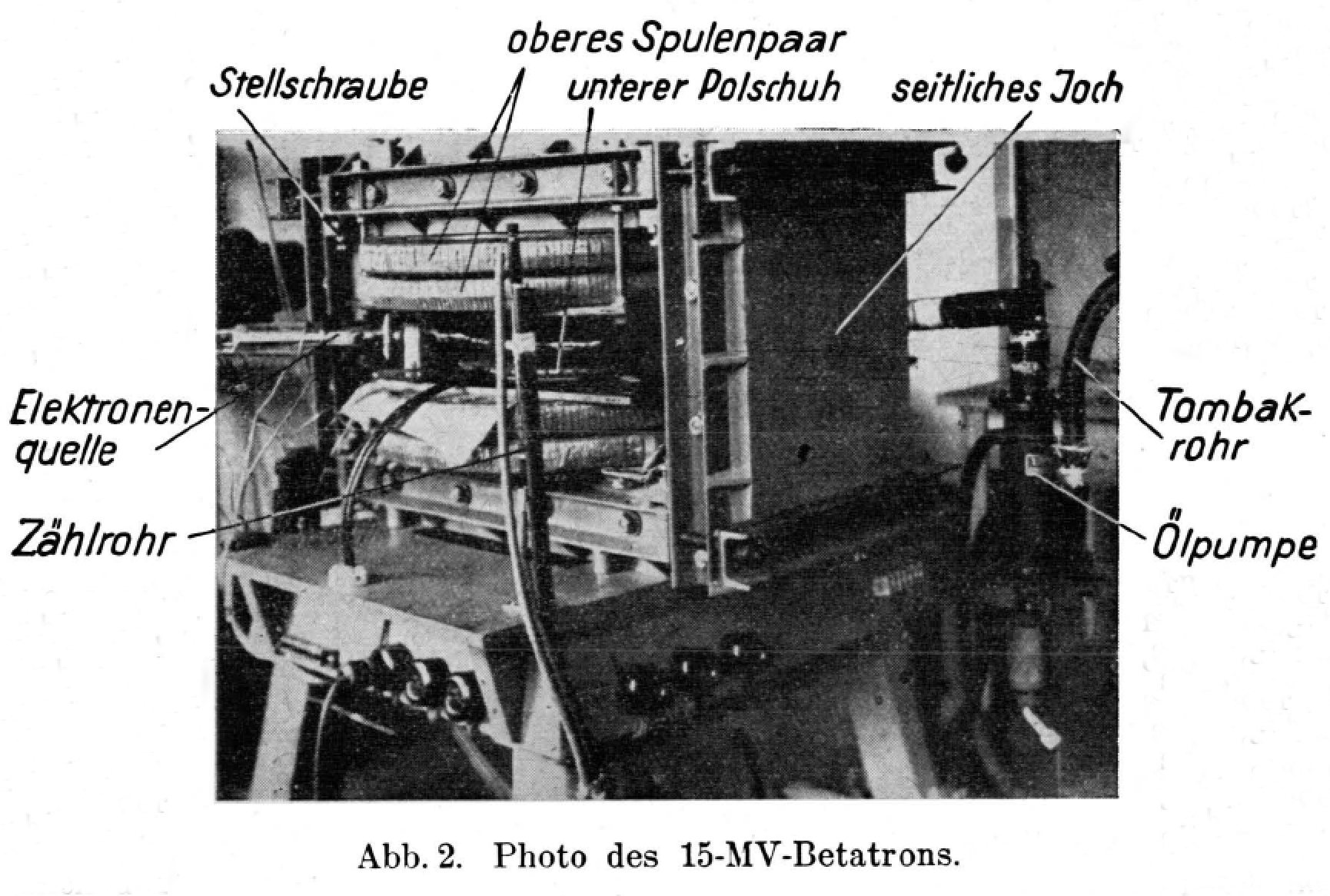}
\includegraphics[scale=0.31]{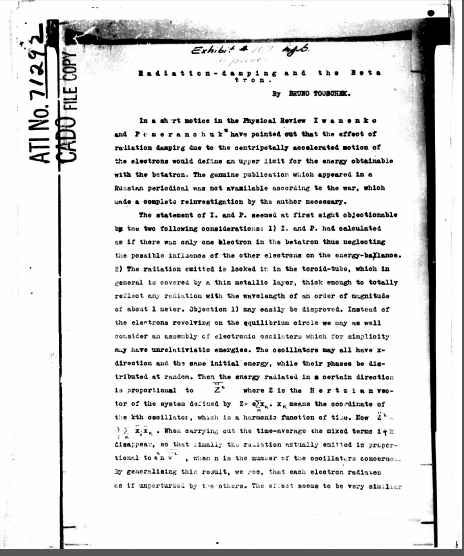}
\caption{At left, Photo of the 15-MeV betatron mounted on a big table as shown in \citep[635]{KollathSchumann:1947}. At right, we show the  first  page of Touschek's 1945 report on Radiation damping in a betatron, unpublished report for the US Armed Forces, 
  part of the Air Technical Index [ATI] collection available at \url{https://apps.dtic.mil/dtic/tr/fulltext/u2/a801166.pdf}. Signed  by \BT \ at the end of the note, the handwritten date reads as  to be 28.9.45.}
\label{fig:betatron-radiation}
\end{figure}

 In October, following British-American careful investigations   held on various German Science and Industrial Institutions, ``investigators'' from the British Intelligence Objective Sub-Committee (B.I.O.S.) visited the C.H.F. M\"uller factory in Hamburg, where the 15-MeV betatron had been built, and where Touschek had worked with Wider\o e and his group. A  photo of the betatron,  from a postwar publication  by two members of the group,    is shown  on the left panel of Fig.~\ref{fig:betatron-radiation}.\footnote{In B.I.O.S. Final Report No. 201, Item No. 1,7, 21, dated 8.10.1945, ``Visit to C.H.C. M\"uller, A.G. R\"ontgenstrasse 24, Bahrenfeld, Hamburg, reported by C.G. LLoyd and G. J. Thiessen, \url{http://www.cdvandt.org/BIOS-201.pdf}, on p. 3  it was further specified   that   ``Dr. Fehr [assistant to Manager] stated that the project had been experimented for  the Luftwaffe with the hope (?) of obtaining a  death ray  for anti-aircraft work." These reports  covered a wide variety of  German scientific and industrial Institutions, and were  authored by officers from B.I.O.S., C.I.O.S. (Combined Intelligence Objectives Sub-Committee) and F.I.A.T (Field Information Agency Technical, United States Group Control for Germany).}  The right panel shows the first page of a report on radiation damping in the betatron, where we see what would become  Touschek's lifelong interest in the question of  how radiation from a moving charge affects the operation of electron accelerators. This report is likely to include  the work   Touschek was working on  during his imprisonment, and which Amaldi mentions as having been written  in invisible ink   \citep[5]{Amaldi:1981} on Heitler's book on the quantum theory of radiation \citep{Heitler}.\footnote{A different version of this work, entitled ``The effect of Radiation-Damping and the Betatron'', undated but bearing an address in G\"ottingen -- and apparently submitted to the  {\it Physical Review} (according to the first line of the document) where it was never published -- is preserved in Bruno Touschek Archive, Box 4, Folder 15.}

The investigation continued with a visit to Wrist, the town near Kellinghusen, where the betatron had been  kept since Touschek and Wider\o e \ had brought it in March 1945. The British officers  were in the Wrist Laboratory, on October 23rd.\footnote{A.T. Starr, K.J.R. Wilkinson, J.D. Craggs, L.W. Mussel, ``German Betatrons'', BIOS, Final Report No. 148, Item No. 1, dated 24.10.1945. \label{bios148}}
In their report it was mentioned that they had received a series of reports written by  Bruno Touschek.\footnote{Copies of such reports on the theory of the betatron written by Touschek (Zur Theorie d Strahlentransformators. Typoskr.-Kopie, o.D. 10 Bl.+ Beil.; On the Starting of Electrons in the Betatron. Kopie, o.D. 11 Bl.;   Die magnetische Linsenstrasse und ihre Anwendung auf den Strahlen-Transformator. Typoskr.kopie, 10 Bl., 1945; Zur Frage der Strahlungsd\"mpfung im Betatron. Typoskr.- kopie, 7 Bl., 1945) can be found in Rolf Wider\o e's papers at the Eidgenossischen technischen Hochscule (ETH) in Zurich (see finding aids at \url{https://www.research-collection.ethz.ch/bitstream/handle/20.500.11850/140811/eth-22301-01.pdf}).} 

Touschek's contribution  is clearly acknowledged also in another detailed B.I.O.S report on {\it European Induction Accelerators} prepared  in October 1945 by the U.S. Naval Technical Mission in Europe: ``In collaboration with the design work of Wider\o e, a considerable amount of work was carried out by Touschek. This is known to have been of invaluable aid in the development of the 15-MeV accelerator. Further theoretical work has also been done by Touschek in the starting of electrons in the accelerator. Some of the work is along the lines initiated by Kerst and Serber which were known to Touschek." It is also specified that ``Wider\o e and the group that came to be associated with him in the war-time German betatron work were not in sympathy with the Nazi-cause, and were persuaded to continue their work for purely scientific considerations."\footnote{B.I.O.S MISC.77, p. 6.} 

The activities of the Megavolt Research Association in Wrist  are examined in detail in the second part of  report No. 148.   The British investigators specified in particular that ``The experimental work at Wrist should close down at once'' and that ``The complete apparatus should be sent to UK''. 
They finally specified that ``Mr. Touschek is recommended to be taken to UK for work on theoretical physics".\footnote{See reference in footnote \ref{bios148}. }
This recommendation sheds light on all of Touschek's whereabouts in  the year to follow. As we shall see, starting in the early months of 1946, Touschek would go back and forth between
Germany and the UK, \Gott \ and London, or Glasgow, until he  would finally settle in Glasgow, in spring 1947, remaining there for 5 years.  

In November, writing to his parents, Touschek  was  still hoping he would soon be allowed to visit  his family in Vienna, in the Soviet occupied zone. 
 But this could not happen yet. The T-force had other plans for him.

In December,\footnote{Letter to his parents,  December 13th, 1945 from Kellinghusen.} working on his research on radiation damping in the Wrist office, from 10 in the morning until 10 in the evening in the only warm room in the area, the uncertainty of the situation was becoming unbearable. The {Western Alliance} were making preparations for the reconstruction of Europe, but in the meanwhile, living conditions were dramatic. Winter was coming, there was scarcity of food, hardly any winter clothing, heating was  a luxury.

As the year was coming to an end, it  was clear that the {\it limbo} in which the German scientists were kept by T-Force 
 could not go on forever, and some decision would and should be taken. 
 Exacerbated, Bruno  wrote to the officers of the T-Force, but no immediate answer came about the decisions yet to take. However,  a  rumor, eventually originated by the T-Force,   gave him the hope    that he would be brought to England. The probability could be low, but the  prospect   made the situation  more tolerable.  Europe, at that time, pillaged of its scientists and infrastructures, was not appealing for his future as a physicist, and, financially,  England would likely be a much better prospect, given that  he wanted  to help his parents, his father being retired from the Austrian Army, and living under difficult conditions in Vienna, under Soviet occupation.
Another reason, the  main one perhaps, was that going to England would make   a plan he had envisioned before the war  come true. In spring 1939, after the Anschluss of  Austria shattered the regular course of his life and studies, while in Rome visiting his maternal aunt AdA Weltmann, 
he had decided to go and  study in England. Actual plans had been drafted and he  had applied for a Visa to the British consulate in Rome, but these plans never materialized.

During these last few months of 1945, the reconstruction of German science was 
being discussed and planned by the occupying forces: how, how much, where, and
under whose direction, these were the questions to pose and  solve.
Finally, the decision was taken
that  post-war German science would be mainly rebuilt in the University town of
G\"ottingen, which had been  relatively untouched by Allied bombing. 
This decision had a
wide impact: the Farm Hall detainees could be allowed to return home, to their families and institutes, and Werner Heisenberg would  be one of the key figures in the revival of scientific research in Western Germany, and especially in German science policy.
Once this path was clear, 
also other decisions came along and the restrictions imposed by the T-Force on the betatron group
were lifted. 

In December 1945,  work with Widero\o e's 15-MeV betatron had been completed at Wrist and the machine was transported to the Woolwich Arsenal near London, where it was used for some time with the help of Rudolph Kollath,  one of the members of Wider\o e's betatron group.
\footnote{Rudolf Kollath wrote a five-pages long report on their results (``Bericht von Ing. R. Kollath, 11.12.1945, \"uber die Arbeiten am Betatron in Wrist'', see copy in Wider\o e's archive in Zurich). Kollath and Schumann, who had operated the betatron  up to the end of 1945, wrote together an extensive report on the performance of the betatron and on tests in Wrist which was published only about two years later \citep{KollathSchumann:1947}. A detailed outline of the 15-MeV betatron and related work carried out by the group, including studies for a large 200-MeV betatron, were reviewed by Herman F. Kaiser in early 1947, also specifying different aspects of Touschek's involvement as a theorist \citep{Kaiser:1947aa}.} 
As for Bruno, he  could be  free to leave Kellinghusen and go back to his studies, the
first thing being to obtain his diploma in physics. 

Bruno had to now start his life
anew. As a 17 year old youth he had gone through the Anschluss, and then the loss of his identity
as a rightful Viennese citizen and  promising physics student,  who had been following courses during Spring 1939
in the University of Rome and, in 1939-1940, at the University of Vienna. He had been living in Germany
through four years of semi-hiding, with little food, scarcely any heat, both in
Berlin and Hamburg with daily bombs devastating the cities, away from his beloved
family, the grandmother, his father, his stepmother, the aunts and uncles from his
large maternal family. At some time, he had learnt that his grandmother
Weltmann  had never returned from Theresienstadt,
the concentration camp 20 kilometers from Prague. The world of his youth was definitely
over. He now had to go on with life. How?
Not unlike many scientists in those days, he could do this only by
fulfilling his dreams. For Bruno, they  were the ones   he had pursued  through his correspondence with
Arnold Sommerfeld in 1941 and which has prompted him to move to Germany in 1942. He had dreamed of studying and 
becoming  a physicist. This is what he  was now anxious to do and was the only way he could overcome the grief for the lost past.  
As we shall see, he was not yet free to decide  his  destiny, and had only a partial  notion of which decisions were taken about the rebuilding of universities in Germany. Likewise, whether he could eventually end up studying  in the UK was also rather nebulous. As things unraveled, his first return to normality was to  be at  the University of G\"ottingen. 

In the section to follow we shall temporarily leave the story of Bruno Touschek, and give a brief overview of what had been  happening in Germany and what Touschek found when he joined the University, in spring 1946.  

\section{From destruction to reconstruction: Starting anew in G\"ottingen}
\label{sec:gott}

Touschek would not be alone in rebuilding  his hopes and dreams. As 1946 started, all around
him the  titanic effort of    the reconstruction of
Europe was already taking place,  coordinated by the American military, with the
UK command on its side. 
{ The reconstruction of science in postwar western Germany -- and German political revival -- is to be framed within the broad contexts of the Allied occupation \citep{cassidy_controlling_1994,cassidy_controlling_1996,judt_technology_1996,judt_denazifying_1996,gimbel_science_1990}.}
As described in Krige's  {\it American Hegemony and the
Postwar Reconstruction of Science in Europe}, ``The immense scientific and technological
achievements in the United States during the war and the ongoing
support for research in the country after 1945 contrasted sharply with
the situation in postwar Europe. There, laboratories were ill-equipped,
destroyed, pillaged, and (in the case of Germany) strictly controlled;
researchers were poor, cold, hungry, and demoralized; and national governments
had far more pressing concerns than scientific (and technological)
reconstruction.'' \citep[1]{Krige:2006}.  However, after the war, ``science had become an affair of state'', strongly intertwined with the re-shaping of socio-economic relations in the wider context of Cold War relations between the United States, the Soviet Union, and the countries of war-ravaged Western Europe. As stressed by Krige, ``Combining scientific advantage with
economic and political leverage, scientific statesmen, officials in the U.S.
administration, and officers in organizations like the Ford and Rockefeller
foundations did more than simply `share' science or `promote'
American values abroad; they tried to to \emph{reconfigure} the European scientific landscape, and to build an Atlantic community with common practices and values under U.S. leadership''  \citep[3]{Krige:2006}.


During the immediate post-war years Germany was facing  devastation, poverty, enormous loss of lives, and  the collapse of economic and political organization with   the country divided  into four occupation zones.  Among the many  challenges,   the reconstruction of research in Germany was extremely difficult, as  German science  had to rebuild itself practically from the ground up and, at the same time, needed to be  reintegrated into the international community. Moreover,
 Allied restrictions specifically forbade applied nuclear physics and in particular also commercial production of betatrons, synchrotrons and all particle accelerators over 1 MeV, including many sorts of equipment. The conditions were perhaps most favorable in the British Zone, where the authorities, especially the liaison officer Colonel Bertie Blount, appeared quite open to a dialogue with German scientists.

In the British zone, the city of \Gott \  had survived World War II without major damage, which meant an invaluable starting advantage for the town and the famous university, the oldest in Germany.\footnote{Named after  its founder, George II, King of Great Britain and  Elector  of Hanover, the Georg-August University of \Gott\ was founded in 1734 with starting  classes in 1737.} With the permission and the encouragement of the British, G\"ottingen grew into one of the main scientific centers of the Western occupation zones.

The Georgia Augusta was the first German university to resume teaching already in September 1945. It had lost its excellence after the great purge of 1933, because of the expulsion from the University or flight abroad   of leading Jewish physicists and mathematicians, among them  
Max Born, James Franck, Edward Teller, Leo Szilard, Eugene Wigner, Richard Courant, Edmund Landau, 
and Emmy Noether.\footnote{A good source of biographical data on eminent mathematicians can be consulted at the site \url{http://www-history.mcs.st-and.ac.uk/} maintained by St. Andrews University.} 

In the early 1930's,  the University included  Institutes for experimental and theoretical physics. The Institute  in Experimental physics  was directed by James Frank, the one in theoretical physics by Max Born. Both had arrived  to \Gott \ as Professors in 1921, and, in due time, both were to win the Nobel Prize.\footnote{The Nobel Prize in Physics 1925 was jointly awarded  to James Franck and Gustav Ludwig Hertz ``for their discovery of the laws governing the impact of an electron upon an atom". Born was awarded the 1954 Nobel Prize in Physics ``for his fundamental research in quantum mechanics, especially for his statistical interpretation of the wavefunction", sharing it with Walther Bothe, as from  \url{https://www.nobelprize.org/prizes/physics/1954/born/facts/}.}
In 1933  the National Socialists' rise to power and the contempt for modern ``Jewish'' physics, which included Quantum mechanics and Einstein's theory of relativity,  forced 
them to emigrate.  James Franck, by then a Nobel laureate, went to the United States. Max Born, one of the founders of quantum mechanics,  went to Italy, then  Cambridge and, in 1936,   to Scotland, at the University of Edinburgh. 
 Born's  chair at the Institute for Theoretical Physics was  then   occupied in 1936  by Richard Becker, who  had been transferred  to \Gott \  by order of the Reich Ministry for Education \citep{Hentschel:2001aa}. Becker and Born had an influence on
  Touschek's  development  as a theoretical physicist, as    Becker was   Touschek's professor when Bruno studied in \Gott\ in 1946. As for Born, Touschek  first met him in Edinburgh,  in May 1947, while in Glasgow as a doctoral student.    Bruno  became a regular attendee of Born's weekly seminars, and, later on,   prepared  the Appendix  on the theory of neutrinos in  Born's new edition of his famous  {\it Atomic Physics}.\footnote{This book  had several English editions starting from 1935, the last one in 1969.}

After the war, the First Institute for Experimental Physics was headed by Robert Richard Pohl, whose research constituted one of the foundations of solid-state physics. The direction of the Second Physics Institute between 1942 and 1953 was in the hands of a former student of James Franck,  Hans Kopfermann  \citep{Weisskopf:1964}, who
 initiated and supported nuclear physics in Germany together with his assistant Wolfgang Paul.\footnote{W. Paul shared  with Hans G. Dehmelt one half of the 1989 Nobel Prize in physics  ``for the development of the ion trap technique", the other half was awarded to Norman F. Ramsey ``for the invention of the separated oscillatory fields method and its use in the hydrogen maser and other atomic clocks".} During the war, Kopferman and Paul  learned of Kerst's success in  constructing and operating the first betatron, and decided to build such an accelerator as soon as possible. The project was put aside as they heard that Konrad Gund had built a 6-MeV betatron at Siemens-Reiniger Company in Erlangen \citep{Waloschek:2012} and Paul started taking measurements  on the machine in Erlangen. After the war,  Paul and Kopferman, with the help of Ronald Fraser, Scientific Advisor of the Research Branch of the British military, were able  to transfer this betatron to  \Gott.  Together  with Becker,  Kopferman was Touschek's advisor for his 1946  Physik-Diploma dissertation about  the betatron, from the University of \Gott\ \citep[7]{Amaldi:1981}.


In early October 1945,  while the University of G\"ottingen was resuming academic activities,
Heisenberg, Hahn and von Laue, while still held in Britain,  had met their British colleagues at the Royal Society in London to discuss the rebuilding of German science.\footnote{In this meeting, two of the German physicists had been Nobel prize winners,  Max von Laue in  1914 and  Werner Heisenberg in 1932.  The third,  Otto Hahn, would be awarded the prize  a few months later, while still under  imprisonment at Farm Hall. On the British side, the meeting included Patrick Blackett, who would be later awarded the  1948 Nobel Prize in physics, for his work on cosmic rays. In early September,  Blackett  and  Heisenberg had held  a long conversation in Farm Hall.  This encounter  had then been  followed by a letter addressed to Blackett by Heisenberg, on behalf of the other scientists. In this letter, the conversations and the position of Heisenberg and the other Farm Hall detainees was summarized. 
 \citep{Bernstein:2001}.}    On 3 January 1946, the ten German nuclear scientists,  among  them  by  now three Nobel Prize laureates, 
were finally released from Farm Hall and  brought back to Western Germany by Colonel Bertie Blount, who had studied in Germany.\footnote{ At his arrival, on January 3, 1946, Heisenberg immediately wrote to his wife Elisabeth from the small village Alswede: ``My dear Li! This is the first evening back in Germany since the end of the war. This long time of captivity seemed to us only bearable through the scientific work. How it's going to be here, we do not know yet. The purpose of our being here is as follows: The highest authorities have decided that we all should in the future have our workplaces in the British occupation zone." \citep{Heisenberg:2016aa}.}
With the British officials Bertie Blount and  Ronald G. J. Fraser, himself a physicist as well as Scientific Advisor of the British military, the group of German physicists started  to  forge working relationships. It was the beginning of a long collaboration which had a great importance for the future of the Federal Republic.

On January 12 Hahn and Heisenberg visited G\"ottingen with Col. Blount, where they found Max Planck, who had arrived there as a refugee seeking shelter with relatives.\footnote{Max Planck had been awarded the 1918 Nobel Prize ``in recognition of the services he rendered to the advancement of Physics by his discovery of energy quanta."} Since 1930, Planck was President of the Kaiser Wilhelm Society, a non-university science organization founded in 1911 to conduct specialized basic research in its own Institutes, predominantly in the natural sciences. The Kaiser Wilhelm Society (Kaiser-Wilhelm-Gesellschaft) had quickly established itself nationally and internationally thanks to its outstanding scientific achievements, but during the 1930s-1940s, the Society's leadership and many of its scientists had become supporters of Hitler's regime, or  had been involved in armament research. The Allies were thus urging that the Society should be dissolved. However, with the support of  Nobel Prize Laureate Max Planck, who was unanimously regarded as an outstanding scientist with an impeccable international reputation,   Otto Hahn's efforts succeeded  in  gaining  British approval for the revival of the Kaiser William Society and on 26 February 1948 the Max Planck Society was eventually founded in G\"ottingen as successor organisation.\footnote{Its first president was the Nobel Prize laureate  Otto Hahn. The Max Planck Society then evolved into one of the mainstays of the science landscape of the Federal Republic of Germany, which was founded in 1949 \citep{walsh_max_1968} \citep{dickson_germanys_1986}.}

Seven of the ten physicists kept at Farm Hall, were now members of the Kaiser Wilhelm Institute for Physics in G\"ottingen: Werner Heisenberg, Max von  Laue, Carl Friedrich von Weizs\"acker, Karl Wirtz, Horst Korsching, Otto Hahn and Erich Bagge. The first four were also given positions as professors at the University. But towards the end of February1946, they had discovered that   the Institute would be hosted in the empty rooms of the former Aerodynamics Research Institute (Aerodynamische Versuchsanstalt,  AVA)  which had been denuded of all its war-related equipment: its large wind tunnels had in fact been partly destroyed and partly dismantled  and transported to England.\footnote{On the race to take possession of  the German aircrafts as well as research and production facilities see \citep{Christopher:2013aa}.} 
All of the scientific machinery and commodities, with which they had hoped to be able to start anew, had been carried away. Moreover, they had no access to all their instruments and equipment  left in  Heisenberg's Institute in Hechingen after the scientists  fled as the Allied army was proceeding through Germany. The Institute  was now located in the French occupation zone, and could not be reached anymore.\footnote{On February 28, Heisenberg wrote his wife: ``Well, here in G\"ottingen things are limping along, more or less. Our rooms in the AVA, at this point, are ugly, some basic office space devoid of any hint of warmth, but useful enough as temporary campsites in the crusade of life.'' \citep{Heisenberg:2016aa}.}

\begin{figure}[htb]
\centering
\includegraphics[scale=0.18]{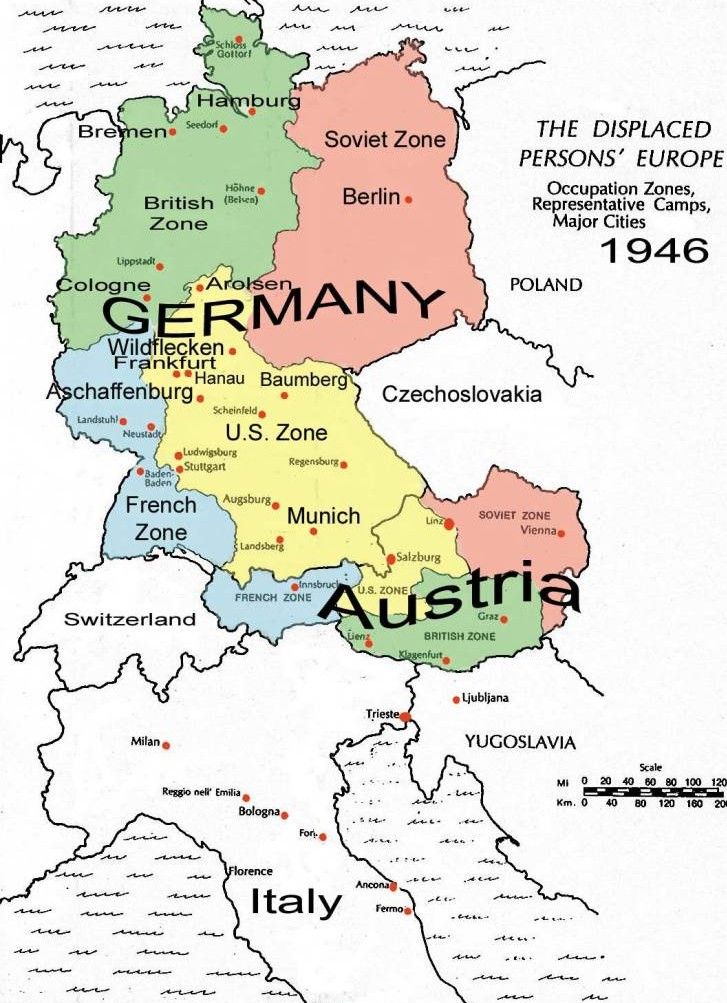}
\includegraphics[scale=0.33]{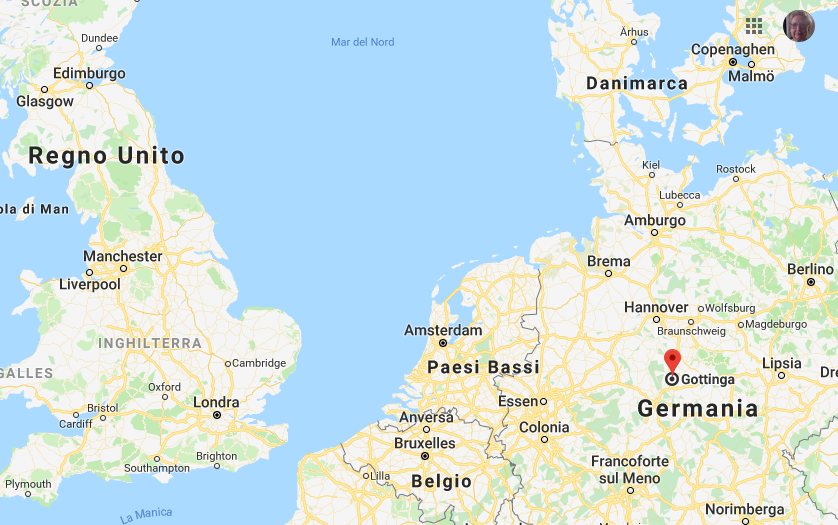}
\caption{At left, we show the four zones into which Germany and Austria  were divided and  became occupied by the winning powers, following the accords  which took place in  Potsdam, a small city located  in occupied Germany, from 17 July to 2 August 1945, map is from \url{https://estonianworld.com/life/remembering-estonias-wwii-refugees/}.  At right an image of the location of the town of \Gott\  in relation    to the places in UK  through which Touschek would later move.}
\label{fig:occupationzones}
\end{figure}

In Fig.~\ref{fig:occupationzones}, we show a map of how  Austria and Germany were divided among the four powers which had won the war, with \Gott \ being in the British zone, some 200 kms North of Cologne.\footnote{A day by day account of how the final agreement about the division among the four powers, can be found at \url{https://berlinexperiences.com/potsdam-conference-1945/}.}

In G\"ottingen, in the years to follow, Heisenberg devoted himself  to two large tasks: the reconstruction of the Kaiser Wilhelm Institut f\"ur Physik as a center for experimental and theoretical research in physics and the renewal of scientific research in Germany, where,
during these early post-war years,  research was  limited by the directives of the Allied Control Commission.

Heisenberg's efforts took large part of his energies, but he was  successful and, 
 during the years of the reconstruction, the Max Planck Institute for Physics  gained a growing reputation as a leading representative of German physics in the international arena after so many years of isolation during the Third Reich and World War II.  As he wrote to his wife in the early days of his \Gott's stay, on February 28: ``What our future will look like in all its reality, I cannot yet tell at all. In spite of it, I have the clear sense that it will not really be all that bad, if only we are patient.'' \citep{Heisenberg:2016aa}.

\section{1946: Touschek between G\"ottingen and Glasgow}
\label{sec:fromGotttoglasglow}

As 1946 started,   Touschek   was anxious  to have a clear  idea of how and where to    continue  and complete his studies. As we shall see, this was not so obvious, and 
through the whole of 1946 he moved back and forth between  the United Kingdom, travelling to London and Glasgow, and  Germany, following courses in G\"ottingen.

In 1946, the decisions leading to the reconstruction of Europe were being put in action. Many European scientists, who had left  to join the Manhattan project, returned to Europe. Together with  those   who had remained in Europe, they could  resume  visiting   each other's laboratories and universities and restart pre-war exchanges. The  first international conferences since he beginning of the war were held.  Everything was slowly starting anew. 

In Germany, science was to be rebuilt starting from G\"ottingen, but all equipment of technical or scientific interest had been taken to England or to the US (as in the case of   the already mentioned wind tunnels, as also discussed in \citep{Jacobsen:2014}). Among them, was Wider\o e's betatron. As Wider\o e writes in his biography: ``In December 1945, the British authorities decided to take the betatron, as part of the booty of war, from Kellinghusen to the Woolwich Arsenal near London.  Apparently, Rudolf Kollath later on took charge of its operation in Woolwich where it was used for non-destructive X-ray inspection of steel plates and such like. The machine has since disappeared without a trace. Many, including myself, later attempted to find it, but
with no success. It was most probably scrapped." \citep[87]{Wideroe:1994}. 

In 1945, to the officers of the T-Force reporting on Wider\o e's betatron, Touschek   had expressed the desire to go the UK, and such had  been the recommendation in the BIOS report. This had been also the  plan he had pursued just before the war broke out  in Europe.
At the same time,  he was also   part of the ``war booty'', to be  interrogated   on `German science', but in particular on the betatron, and, sometime in January or February 1946, he    was taken to England.\footnote{This early visit to London is glimpsed from  a letter he sent to his parents  on April 8th, from Glasgow, where he  mentions that his entrance to the U.K. had been again refused, an indication  that he had already been in  the UK, but that refusal of entry did not prevent him to enter the country. The apparent contradiction between  immigration authorities and the military, which were accompanying Touschek as an `enemy alien', is similar to  the  long drawn fight between the US  Immigration and Naturalization Service and military  authorities over allowing entry or residence rights to German scientists, who could have been  involved in war crimes, as seen in \citep{Jacobsen:2014}.}
During this first visit, plans for Touschek's move to study in a university in the UK were put in motion.
The English officer  in charge of the  German scientists in G\"ottingen was Ronald Fraser, whom Heisenberg remembers as a friendly British officer in his memoirs of the period.\footnote{Ronald Fraser was a research physical chemist at Cambridge University, where he worked for a few years after having been a lecturer at Aberdeen University, as from footnote (50) in  \citep{Amaldi:1981}.}  Fraser  sought to have Touschek to go to the University of Glasgow, where plans for  300-MeV electron synchrotron,  to be built under the direction of Philip  I. Dee, were  considered, together with a smaller,  preliminary 30-MeV machine.\footnote{ Fraser  knew Dee from  the University of Cambridge, where Dee had graduated in 1926,  later working at  Cavendish Laboratory.}

 Philip Dee had arrived in Glasgow in 1945 to occupy the  long established Chair of Natural Philosophy, which had been offered to him already during the war, while he was involved in radar work and other leading war activities  \citep{Curran:1984aa}. He had immediately set up major plans for relaunching physics, which included the building of an electron synchrotron, based on the new revolutionary principle of phase stability just discovered, simultaneously, both in the USSR \citep{Veksler:1946aa} and US  \citep{McMillan:1945aa}.\footnote{After the war, four types of accelerators were in use: Van de Graaff, Cockcroft-Walton, cyclotron, and betatron. The cyclotrons, which were able to produce the highest energies, had reached their energy limit due to the relativistic mass increase at very high particle velocities, laying at roughly 25 MeV for protons. The principle of phase stability came as a solution to this problem, making it possible to accelerate particles into the GeV region compensating for the relativistic mass increase either by changing the accelerating high-frequency voltage or the magnetic field strength during the acceleration of the particles. Not only cyclotrons could be operated at higher energies converting them into synchro-cyclotrons, but it was also possible to build a completely new type of accelerator, the {\it synchrotron}. This new machine could keep the particles on a path of constant radius by varying both the magnetic field strength and the frequency of the accelerating voltage with increasing particle energy. Last but not least, this kind of accelerator could be used for accelerating {\it both protons and electrons}. Machines based on this principle promised to displace betatrons as accelerators of high-energy electrons: indeed further developments of the betatron  mostly took place  for  medical uses. In US, the leading country in the field, accelerator programs for nuclear physics research were being carried out at Brookhaven and Berkeley, two Laboratories  which played a role as models for European physicists. In fall 1946 Lawrence's 184-inch synchro-cyclotron was producing its first beam at Berkeley's Radiation Laboratory and new machines were being planned, notably a 10 GeV proton synchrotron.} 
 At that time, the only European country, whose scientific and technical evolution in nuclear and atomic physics  could be compared to that of the US, was Great Britain, as underlined by John Krige \citep[488]{Krige:1989aa}: ``[\dots] as the leading nuclear power in (western) Europe at the time, Britain alone amongst European countries had the human and financial resources, and the political will, to launch a major accelerator construction programme immediately after the war.'' 
  
  The foundation of Britain's post-war accelerator construction program were laid out,  immediately after Japan's surrender in August 1945, through   a government committee (Cabinet Advisory Committee on Atomic Energy)  which should advise the new Labour Prime Minister Clement Attlee on general policy for Britain's postwar atomic program. A Nuclear Physics Subcommittee was created on October 4, 1945. It was chaired by James Chadwick and was composed of leading nuclear physicists such as Patrick Blackett, John Cockcroft, Charles G. Darwin, Philip Dee, Norman Feather, Mark Oliphant and George Thomson. One of its first recommendations had been that ``immediate support be given to Oliphant's and Dee's proposals to build accelerators at Birmingham and Glasgow universities, respectively.'' \citep[488-490]{Krige:1989aa}.\footnote{For a discussion of the British projects on accelerators see  \citep{Mersits:1987aa}, especially Section 1.3.3. As part of the British nuclear-physics program, a variety of different types of accelerators was being also planned at Harwell, the site chosen for the Atomic Energy Research Establishment  to cover all aspects of the use of atomic energy, but this program was more oriented towards nuclear physics rather than ``meson physics'', as nuclear and particle physics was called at the time. When the 400-MeV synchrotron Liverpool machine went into operation in 1954, it was Europe's biggest synchro-cyclotron until 1957, when the CERN 600-MeV synchro-cyclotron was completed. Three of these UK university accelerators under construction would allow to do meson physics. Even if in the meantime higher energies had become available in US, they were a good basis for launching a research program in particle physics.}
 Dee's plans were thus part of larger program launched in UK universities between October 1945 and March 1946 which included the building of  big accelerators in five universities: 1.3 GeV proton synchrotron (Birmingham), a 400-MeV synchrocyclotron (Liverpool), the 300-MeV electron synchrotron in Glasgow, as well as two less powerful machines in Oxford and Cambridge.

In this perspective, Touschek's experience with Wider\o e's betatron would be an important asset for Dee's department and the foreseen project. As mentioned, there is some evidence that Touschek was brought to the UK  in the early months of the year 1946, probably to be further interrogated about the German betatron and start  negotiations for a   move to  Glasgow.
A suitable salary   from the  Darwin Panel Scheme, under which German scientists and technicians could be  employed in UK, may  have    been  discussed  at the time.\footnote{Information about the Darwin fellowships,  and the scientists whose  work in the UK was sponsored through the Scheme, 
is  available at the UK National Archives, \url{https://discovery.nationalarchives.gov.uk/details/r/C258396}.}

To finalize such an appointment, it was necessary to wait for the UK government's final approval of the Committee recommendations about the construction of new accelerators.  In the meanwhile, 
Touschek,  still under the `protection' of the T-Force,
was brought  back to  Germany, firstly  to 
  Kellinghusen, to take leave of his apartment and pack  his few things.
As for the next step, while waiting for the Glasgow situation to become definite, the natural choice  was for him to go to  G\"ottingen,  where the University was restarting in the British occupied zone. 
 Of interest to Bruno, was also that  Wolfgang Paul, Kopfermann's assistant at the Institute for Physics of the University, was working with the betatron built by Konrad Gund for Siemens in Erlangen \citep{Waloschek:2012} and which later was brought to G\"ottingen. This would  give  Bruno a good opportunity for discussions with Paul and Kopfermann while he was completing his dissertation to earn his Diploma in physics, the pre-requisite for any further studies.



\begin{figure}[htb]
\centering
\includegraphics[scale=0.83]{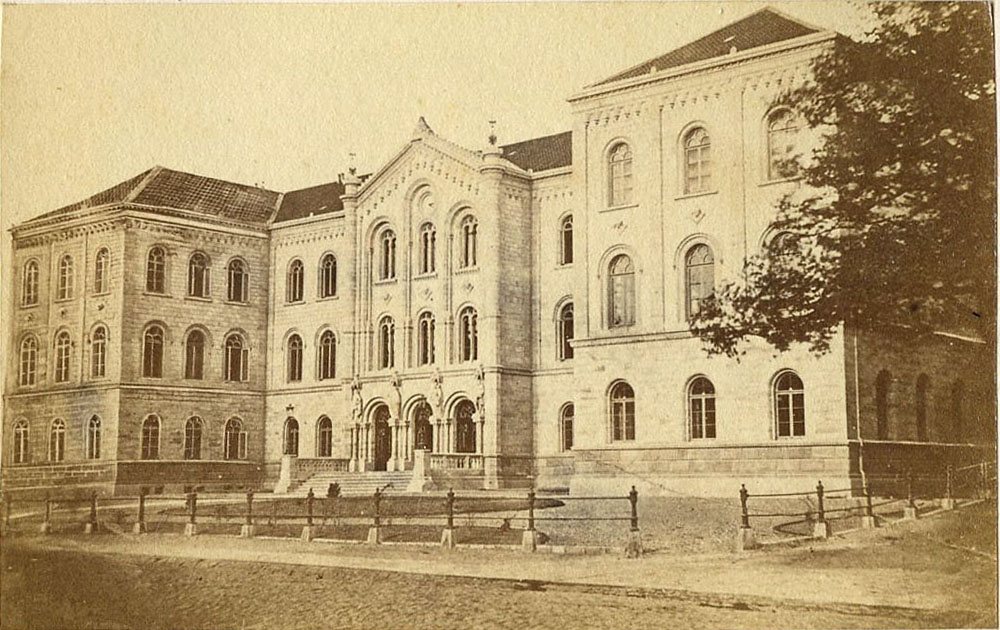}
\includegraphics[scale=0.52]{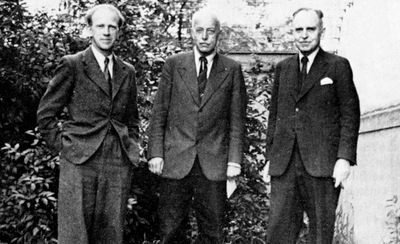}
\caption{At  left,  an image  of G\"ottingen University. At right,  Werner Heisenberg, Max von Laue and Otto Hahn in G\"ottingen, Germany, in 1946. Image: INTERFOTO/AKG-IMAGES, from \url{https://www.nature.com/articles/503466a}.}
\label{fig:heisenbergvonLaueHahn}
\end{figure}
 
Sometime in early March, Bruno  moved to  \Gott \ from Kellinghusen.\footnote{\BT's letter, in which he describes his first impressions of \Gott,  is  dated  12.4.1946, but    the month, as  written,  is   likely to be an error, with Touschek  typing a 4 (April) instead of 3 (March). All  evidence from the letters of this period, in particular   two letters from Glasgow,  respectively on April 8th and 12th,     points to the date ``12.4.1946" to be  ``12.3.1946".} 
  The three Nobel Prize laureates from Farm Hall,  Werner Heisenberg, Max von Laue and Otto  Hahn  were already there, having arrived   since January. We show them in Fig.~\ref{fig:heisenbergvonLaueHahn} together with an old image  of the University, which had been left almost untouched by the war. 
 Touschek knew the group of German physicists   he saw as he arrived in G\"ottingen, since his Berlin days. 
 
\begin{figure}[htb]
\centering
\includegraphics[scale=0.06]{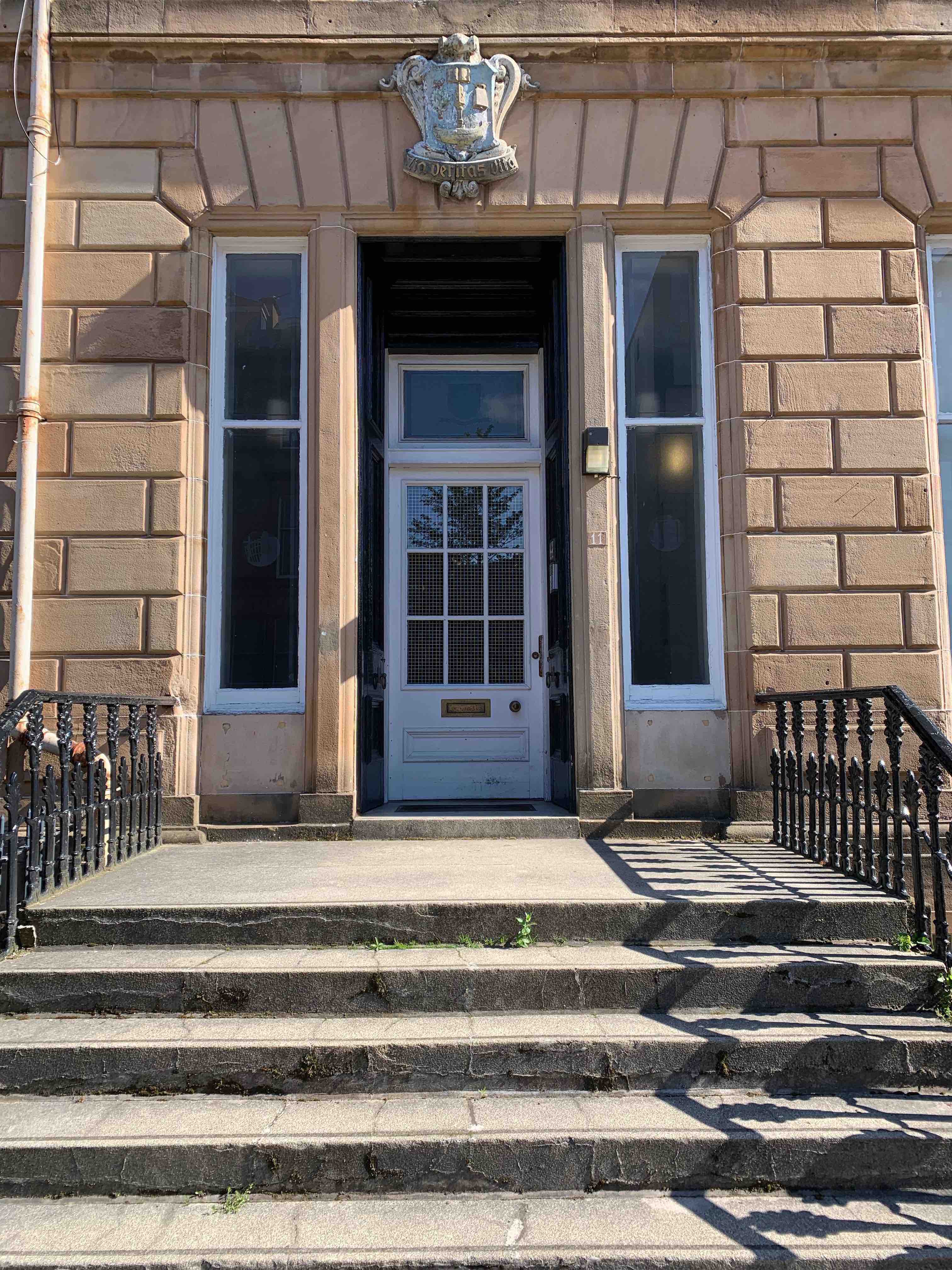}
\includegraphics[scale=0.06]{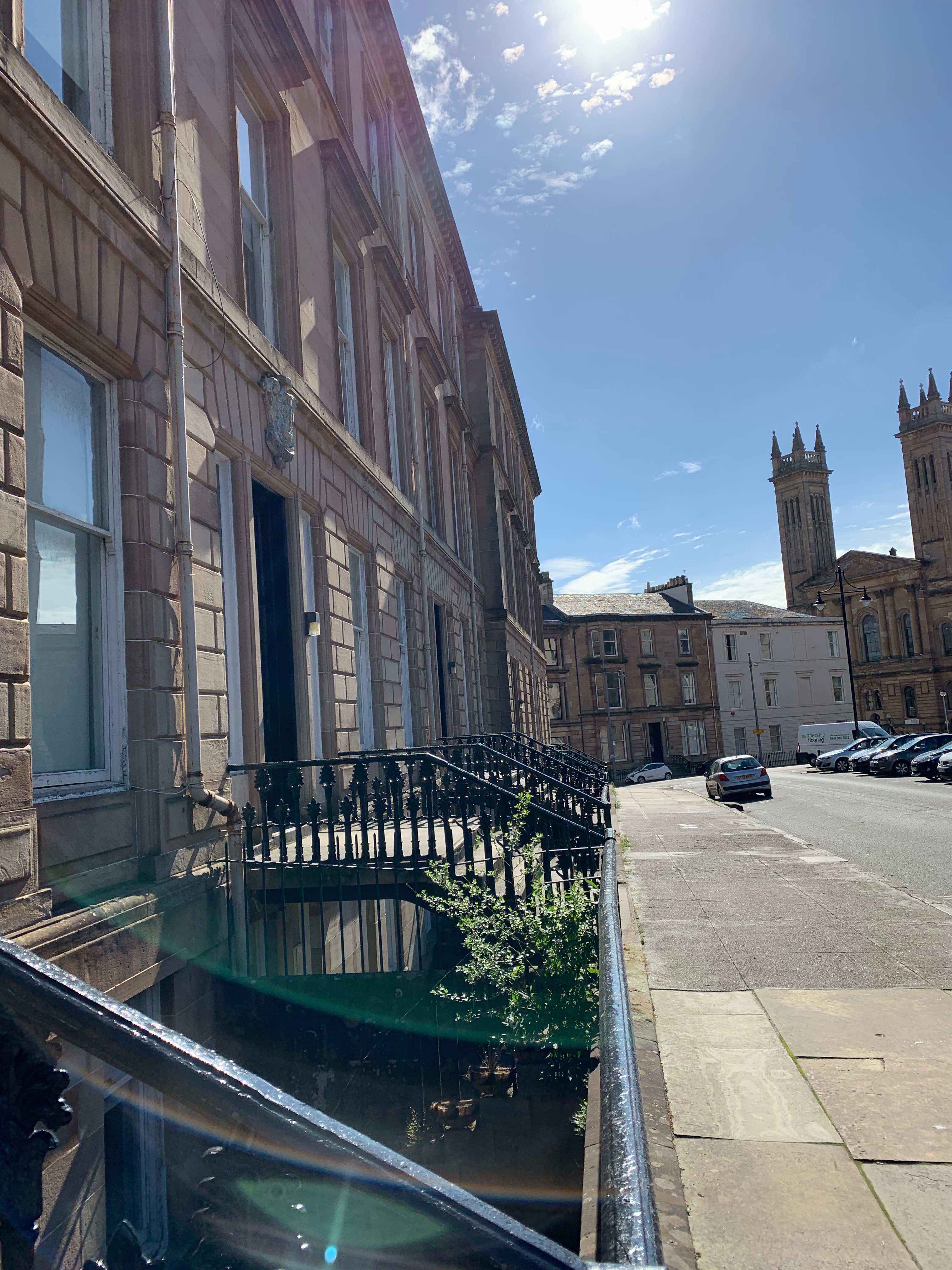}
\caption{Two contemporary photographs of 
McBrayne Hall of the University of Glasgow,  a  front view in left panel, and a  side view   in the right  panel,  courtesy of Dr. Robert McLaughlan.}
\label{fig:glasgow-hall}
\end{figure}

Touschek also renewed his acquaintance with Fritz Houtermans, who was planning to go to Vienna in the summer, raising Bruno's hopes to go with him and finally be reunited with his family. But he was not yet free to do as he wished. Unbeknownst to him, the British plans for post-war accelerator physics development  were being finalized, with approval of the construction of the 300-MeV synchrotron in Glasgow. The plan to bring Touschek to the UK had also been carried through, with arrangements with Philip Dee, and a suitable salary higher than could be expected in Germany or Austria. 
 A contract was prepared for a six month position under the Darwin Scheme  and in April 1st Touschek  was brought  to Glasgow, and housed in 
  MacBrayne Hall of the University of Glasgow, which we show in two contemporary photographs in Fig.~\ref{fig:glasgow-hall}.\footnote{April 8th, 1946, letter to parents  from Glasgow.} 
The accomodation in the old Scottish University   was very different from what he had  seen in \Gott, where he had been housed in the  buildings of the AVA (the Aviation Institute), which had been deprived of its instrumentation, but had been newly refurbished, all shiny and polished.  In McBrayne Hall, his rooms were small and ancient looking, only  1.80  meters high, with beam ceilings running as if towards some distant adventure,  one room   filled with lumber and a bookshelf, crossed by the tubes of all bathrooms, which Bruno  decided, right away,   to paint   pink as  soon as possible. He expected to be there for half a year, at least, and  had brought  his books and few things, which he immediately arranged around the room.  Lack of proper clothing was still a worry, but Bruno had an uncle from the maternal side, Alfred Weltmann, who lived in Birmingham, and whom he was planning to  apply to in case of need.

However, once more, things were not to go on as he had expected. A week since his arrival, a complication arose.\footnote{April 12th, 1946, letter to parents, from Glasgow.} As it  turned out, this was not a small mishap, instead it was a    tough obstacle to overcome. This was so because the  Department for Scientific and Industrial Research (D.S.I.R.)   had suddenly found out that he was Austrian, whereas  the Darwin panel, from which Touschek's salary    should come,  only applied to Germans. Although this had been clearly stated by Bruno in the many questionnaires which he had  filled  in the intervening months, this `detail` had obviously escaped the attention of whoever had prepared the contract.

 This  was an unexpected drawback. Once more Bruno's   path  had to be changed. Somewhat used to skirt  administrative regulations,  Touschek at first thought the objections could be of minor import and  underestimated the difficulty of overcoming the D.S.I.R. objections. Since leaving Austria as a twenty one year old, Bruno had either been in semi-hiding or under the control of military authorities,  while working in Wider\o e's group during the war, or under T-Force authority  afterwards.  
Thus, he did not understand that, outside the control of the  military,   civilian life was quite differently regulated  and administrative obstacles were  not so easily  overcome.   
All his life, Touschek would have little tolerance for this type of delays and   encumbrances, also  a remnant of how, in the war years, he had to find alternative solutions to survive. As ultimate instance, one can remember that, towards the end of his life, he refused to prepare and submit  his scientific credentials for promotion to Professor of Physics at the University of Rome \citep{Amaldi:1981}. His friends had to do it for him. 

\subsection{Getting a Diploma in \Gott}
The D.S.I.R. proved to be a hard contender and  very soon 
Touschek  left to  return to G\"ottingen, to prepare for the  physics exams and presentations leading to his Physics Diploma, while, in Glasgow,  Dee would continue his efforts to have  him  join the University.

Touschek  was not disappointed about having to return to \Gott.
He was very resilient: a life of losses and changes, starting with losing his mother when in his early teens, then the expulsion from Vienna University and the years in Germany  spent almost in hiding, up to miraculously escaping death during the final days of the war, all this had hardened his resolve to survive and bounce back.  He was still young and confident in his future. 

Back in \Gott, he was  pleased  to be with  familiar faces  and have  nice   arrangements  for housing in the countryside.\footnote{May 9th, 1946 letter to  parents from \Gott.} He had immersed himself in  his studies, 
to prepare  the  exams for the Physics Diploma at \Gott. He  passed his pre-diploma exam very well  on May 8th. There also  appeared a thrilling prospect, namely  that, after the exam, he could remain,  for a while at least,  as a research assistant with Heisenberg's group  in particle theory. This   was  a chance like he had never encountered  before and could not be missed, after  almost seven  years of disrupted life. Thus, the plan to go to Vienna in the summer  had to be postponed, notwithstanding his parents' pressing for his return there.

In fact, having seen Bruno return to Germany from Glasgow, his parents had started hoping he would come back to Vienna.  Earlier he had also received  an offer  for a lectureship in Berlin, and   similar chances could possibly exist in Vienna, as well, but none of  this was in Bruno's plans. He refused the Berlin offer to lecture in electricity and theoretical nuclear physics, because, at that point in his life, the priority for him was to become {\it a physicist}.  A precarious and temporary position in a University, which at the time was completely empty, had no interest for him now that physicists such as von Laue and Heisenberg  were no more teaching there.

About his return to Vienna, this had to be postponed at least by another year, as he  could not see himself going back without  having first gotten his degree and become Heisenberg's assistant in G\"ottingen, an  extraordinary opportunity opening up for him in the coming months.   

\begin{figure}[htb]
\centering
\includegraphics[scale=0.638]{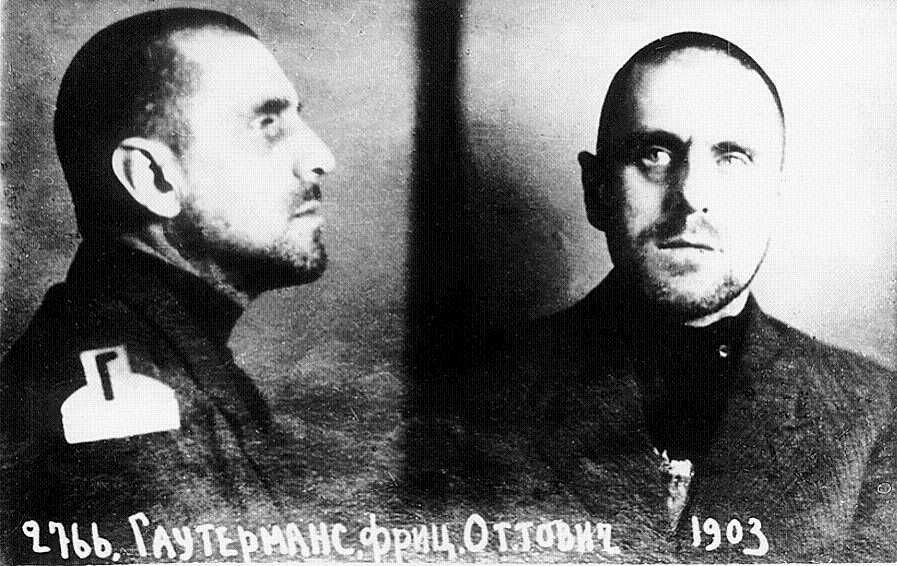}
\vspace{0.5cm}\includegraphics[scale=0.25]{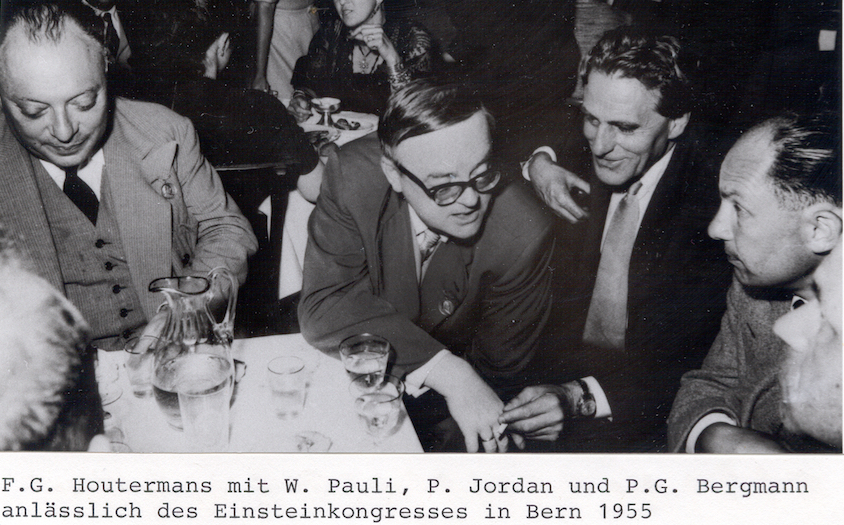}
\caption{ A 1937  photo of Fritz Houtermans while imprisoned by  the Secret Police of the Soviets, from \citep{Frenkel:2011aa}. At right, from left, Wolfgang Pauli,  Pasqual Jordan, Fritz Houtermans and  the well known relativist Peter G. Bergmann during the conference held in Bern in 1955 to celebrate the 50th anniversary of the formulation of special relativity by Einstein,  photo property of the  University of Bern collection, courtesy of S. Braccini.}
\label{fig:houtermans}
\end{figure}

Now that he was engaged in a clear path for his Diploma, Bruno  could enjoy the friendship of other Viennese physicists or professors he had seen in Berlin or Hamburg during the war. One such occasion was on May 10th, when  there was an evening at the Houtermans' to celebrate  Touschek's pre-diploma  exam, with three  Generations  of Viennese scientists: the 75 year old mathematician  Gustav Herglotz,  Fritz Houtermans, then 50 year old,  and  Touschek himself, at 25.  During these months in \Gott \  Touschek became close to Fritz Houtermans. They had both been born in Vienna, came from similarly assimilated  Jewish families, and had both experienced a  skirmish with  death, to which they had come close but from which both had luckily escaped from.  In Fig.~\ref{fig:houtermans}, we show two photographs of  Fritz Houtermans, with the one  in the left panel from the time when he was held  prisoner  at  the Lubyanka in Moskow, in 1937. In the  right panel, Fritz Houtermans is second from the right. 

This time of Touschek's life reflects a  close camaraderie with other German or Viennese physicists, who had all lived through the hardship of war. 

On June 9th, 1946,   Pentecost, {also known as Whitsunday in the English world and  }an important  Christian  festivity, took place.\footnote{June 14th,  1946, letter to parents from \Gott.} The first Pentecost after the war had ended, it held a special importance in Europe. After 
the carnage, divisions and  conflicts of WWII, survivors and warring armies were sharing the hopes and burden of reconstruction, in a kind of suspended peace, which  would soon be shattered by new divisions brought about  by the {\it cold} war. But on that Pentecost Sunday, it was a good moment, for all,   to celebrate the  peace, no matter what their  religion was. 

Touschek,  while studying hard for his exams and presentations, and  basically on the eve  of his diploma preparations,  was one of many other Europeans who shared  this holiday with friends, taking a small break from everyday occupations.

The week-end was quite exhausting. On the  Friday before Pentecost, Houtermans had to go to the observatory, and asked Touschek to follow him there  with Fritz's wife, promising  visions  of a comet. But, first they failed to  find the observatory,  wandering  around until 11 in the surrounding forest, and then,  when they  finally got there, there was no sign of the promised comet, and  they had to content themselves with  a flickering Jupiter. At 2 o'clock they were  not yet back home and the night was lost. 

\begin{figure}
\centering
\includegraphics[scale=1.75]{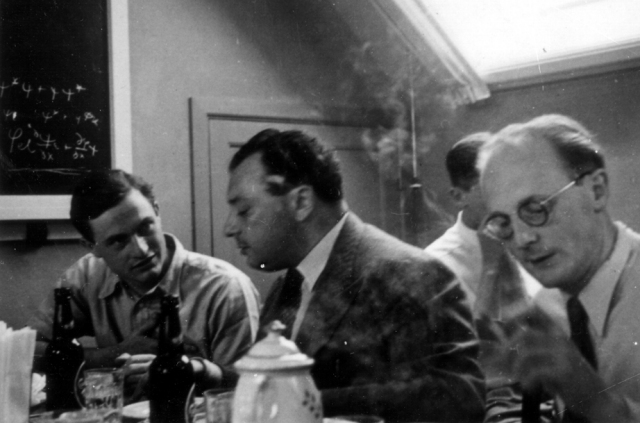}
\includegraphics[scale=0.5]{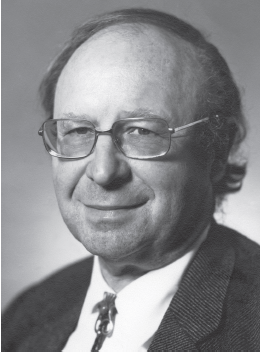}
\caption{Left panel: a photograph of Hans Jensen, at right,   with Wolfgang Pauli, center,  and Markus Fierz (Pauli Archive Photos, CERN, \url{https://cds.cern.ch/record/42961?ln=en}).
 Right  panel: Hans S\"uss from the Biographical Memoirs of the US National Academy of Sciences \citep{Waenke:2005aa}.}
\label{fig:jensen-suess}
\end{figure}

On Saturday,   Jensen from Hanover appeared and they went together to   Houtermans' home, to find  Heisenberg,  a Military Government  official,  and  S\"uss, with whom Bruno  had often come together  in his small apartment in Hamburg, during the war.\footnote{Hans  S\"uss studied  physical  chemistry  at  the  University  of Vienna, receiving   his  Ph.D.  in  1936.  He was in Hamburg at  the Institute   for Physical Chemistry, since 1938. He had a wide range of interests, and becamed an expert in  heavy water, becoming  a  scientific  advisor  to  NorskHydro, the Norwegian plant in Vemork. After the war, in 1950 he  moved to the US. For details see \citep{Waenke:2005aa}.} Saturday night was spent at Touschek's place, with a 'little night physics'  (`eine Kleine Nachtphysik'), in a typical  Jensen-Houtermans' meeting.\footnote{Houtermans was famous for his hospitality. In \citep[27]{Amaldi:2012}, Houtermans' first wife, Charlotte Riefenstahl, is quoted as remembering that in Berlin, around 1930, their  ``\dots  small house and the tiny garden were always bursting with guests. It was not unusual to have 35 people dropping in for tea''. One evening almost every week, the Houtermans invited their colleagues and friends to what Fissel called `Eine kleine Nachtphysik' paraphrasing Mozart's `Eine kleine Nachtmusik'. During these evening get-togethers, discussions around physics often lasted for hours and until late into the night. See also preface in \citep{Rossler:2007aa}.} 
The rest of the night was not much fun, given the rather crampled accomodation in Touschek's place (such as just one bed and both Touschek and Jensen having to share it).
In Fig.~\ref{fig:jensen-suess} we show two (later times) photographs of both Jensen and S\"uss.

On  Sunday morning,
  they  were again all together at Houtermans and Heisenberg was there as well, but apart from Heisenberg, they were all quite sleepy. The conversation must have been sleepy as well, given that, for some of them, the hours from 1 to 5 in the morning had gone by with the `kleine Nachtphysik'.
While all this happened, Touschek was very worried about preparing a lecture he  was to hold  at Heisenberg's seminar, and, in the same days,  completing the submission of his diploma thesis. He was able to bring both to a successful completion, not without obvious effort and strain, and  he   submitted the thesis on June 14th, as he proudly announced to his parents.\footnote{June 14th, 1946, letter to parents from G\"ottingen.} The diploma thesis  on the theory of the betatron had been done under the joint supervision of Richard Becker and Hans Kopferman and was {most}  likely  based on the work he had done during the war and the reports he had prepared afterwards for the T-Force. As for the lecture,  it was  received well and Bruno  could rest and relax for a few days.


In the meanwhile, two weeks before, the {possibility}  of  going   to England had come up again. He had to fill a rather long  questionnaire with an English officer  (taking him well of two hours),  about  a still  rather uncertain 
stay   for a six month period. Among a number of different opportunities, the UK option was still   appealing to him, partly  as he  felt  he owed the British a lot. There were   also   an invitation from \RW\ to visit Switzerland for a three week period, and the offer for a  lectureship position at the Berlin University, which he had already decided to refuse. In any case,  nothing could be decided  until his diploma had been granted.

His gymnasium papers, testifying that he had passed the {\it matura} in Vienna in 1939 at the Staattsymnasium, were requested and received, and he passed his Colloquium with full honors on  June 26th.\footnote{June 28th, 1946, letter to parents from \Gott.}
 At this point, after six months of having  gone back and forth between London, \Gott \ and Glasgow, he started asking  what could he do next, or, rather, where would he go.
Beyond  the six-month position under Heisenberg, the plans for the future
 included  the project in Scotland, the Swiss offer, or remaining  in \Gott \  and starve. 
Each of the plans had its own attraction, and staying in \Gott\ with Heisenberg was most appealing to him scientifically.  Financially, however it was   the least secure, because of the lack of research funds available  in Germany at the time. Touschek wanted to help his parents in Vienna, where conditions under the Soviet occupation were very harsh, and the difficulty of doing this, as  a poorly rewarded Heisenberg's assistant,  were scarce.
Waiting   for the work  at the Heisenberg institute to start in August,  
he envisioned to take a small break, such as  driving around the countryside, something  he would  enjoy, but  seemed frivolous.  As a matter of fact,  the decisions he was agonizing about were not in his power to take.

The British in fact had been  preparing his next visits to London, the first of which  took place in early July, but in June he would not know about this, and became restless. One night, in late June, after his diploma, he read a book which drove him to reconsider  what had happened in Germany. 
The book was  {\it Darkness at  Noon} by Arthur Koestler.\footnote{Arthur Koestler (1906-1983) was  born in  Budapest, from Jewish parents, who left Hungary for Vienna in the 1920's.   He became a  member of the German Communist party in 1931 and  traveled to Russia. He was disillusioned by what he saw, and, after many perilous adventures which included   Spanish prisons in 1937 and a  stint with the French Foreign Legion, he went to England, and was  later naturalized a British citizen. {\it Darkness at noon} was published in 1940.} It dealt with the fate of a People's Commissar during one of the Russian purges, started in 1933-34, and leads the reader  from the arrest to the hanging. Apart from leaving him quite depressed, he was led to consider the difference between what had happened in Germany and the Soviet still ongoing brutality. He could clearly see how things had now changed in Germany, at least in the English zone. He also saw that people around him did not realize this change, neither the British, nor the  Germans, who had not seen evil when it was in front of their eyes during the Nazi regime, and could now hardly wait for it to become history.

\subsection{Doubts and uncertainty}
\label{ssc:doubts}
 Shortly after receiving his Diploma, Bruno was taken to England.\footnote{July 3rd, 1946 
 letter to parents from London-Wimbledon.} 
In Wimbledon, at   Beltane   School   in Queensmere   Road,  there was an internment camp, where German scientists and technologists  were held in order to obtain information  and  expertise  by interrogating them about techniques in which  Germany was ahead of Britain \citep{Gimbel:1990ab}.\footnote{`Once the Germans [scientists] had been located by the search teams, escorting officers were detailed to accompany them to London where they were taken to an interrogation center in Wimbledon, based at the premises of the Beltane school.', in  \citep{Longden:2009aa}. The center was removed to Hampstead in 1947.} 
Unlike others, Touschek was actually free to move in and out of the Beltane school and was even financially compensated.
By July 19th, he  was   back in G\"ottingen, although not for long.\footnote{July 19th, 1946, letter to parents from \Gott.} 
The frequent moves between the UK and Germany, which appear to have taken place  between July and September,  compounded Touschek's feelings of displacement, even affecting the research he was engaging in. We see that,  for Touschek, the period after his diploma  became a period of great uncertainty.

After the diploma,  Touschek was offered a six month assistantship in G\"ottingen and  he seems to have entertained various possibilities for his future studies, including
to remain in Germany, perhaps doing his doctorate with Heisenberg. Envisaging  the possibility  of a doctorate under Heisenberg was shaking his original desire to go to England. In any case, he 
 now faced two possible pathways to follow,    whether to remain in Germany  for his doctorate, either in G\"ottingen or perhaps in Berlin, or pursue the UK road, to Glasgow, where Philip Dee was continuing  his efforts to obtain for him a doctoral stipend. Both personal and financial reasons weighted in, pulling him in one or the other direction,  and would make Bruno  alternate between different  routes. 

 It was an extremely difficult choice. In Germany, he  could have  the chance to work with Heisenberg, and be surrounded by the top German physicists, eager to rebuild the pre-war eminence of German science. From a strictly scientific point of view, however, there were strong restrictions by the part of the occupying forces   on the  research topics which could be pursued by German physicists. Certainly no accelerator could be built in Germany, for quite some time. And,  soon   to happen,  as we know now, {\it a posteriori}, the greatest advantages in theoretical physics, the development of relativistic quantum field theories and Quantum Electrodynamics, 
  would in fact take place away from Europe \citep{Schweber:1994aa}.\footnote{
  In 1965, the Nobel Prize in physics was assigned to Richard Feynman, Julian Schwinger and Sin-Itiro  Tomonaga for `their fundamental work in quantum electrodynamics, with deep-ploughing consequences for the physics of elementary particles.`} 
On a personal basis, while he had known many of the G\"ottigen professors, who held him in good consideration,   he was an Austrian, would still be partly an outsider, and his Jewish heritage clashed with remaining in Germany. He would of course be even more of an outsider in the UK, where he would be an ex-enemy alien,  but he also had family in Birmingham, a maternal uncle, Alfred Weltmann, with whom he could relate. Ultimately, Touschek always remained an outsider,  and this may have been both the source of his genius, and his demise. 

However, 
it is not clear at this point how and why he followed the original plan and left Germany for Glasgow. 
As we shall see, at the end,  after literally going back and forth between the UK  and G\"ottingen from July to December,  
in April 1947 he   moved to Glasgow. And from Glasgow, to Rome, where he would be the moving force  for   the early development of particle colliders: {\it a  posteriori} one can say that this turned out to be the right choice. 

In the uncertainty, he went back to physics, to a neutrino physics problem he had worked on before. Having  lost all his notes because of  his many moves, and not  able to reconstruct right away the arguments and the calculations,   he felt like an old man, losing his capacities. He even doubted of  losing his talent. 

As for  his future, conversations with Ronald Fraser  did not help to clarify his mind or what could he expect to happen. Fraser wanted to know about possible  work and publications on the betatron,
 but Touschek was now almost totally disinterested in  anything  connected with that work.  Albeit late at this point, Fraser also   gave him  a gratifying information, namely that things were no longer secret and that the whole secrecy  about   the betatron, as it  was in the previous November and December,  was an invention of subaltern officials.  
 
In the second part of July,  while in \Gott, discussing with Fraser whether   Touschek were  free  to accept {a possible}  offer to go to Glasgow, 
 a British corporal appeared with a telegram from England requesting {once more} the completion 
  of {yet} another questionnaire. All this was {still} non-committal,  and Bruno 
   was feeling more and more displaced and without a safe direction to go. Memories of his family were coming back to him  more often, and,  at times, he  dreamed of  taking  a vacation, three years from now, after his doctorate,  and go back to the `Colle d'oro', the golden hill near Rome, where his aunt Ada had a summer house and where she had taken him, during his visits before the war.\footnote{The `Colle d'oro' is  a location near Velletri, one of the many small towns dotting  the vulcanic hills South-East of Rome.}

Between August  and December, Touschek was in the UK at least one more time.\footnote{August 18th, 1946, letter to parents from Wimbledon.}
In August he may also have been again in Scotland, where the  position 
in  Glasgow University, 
in the department where Lord Kelvin had held a chair,   was appealing and definite enough 
that he gave up his room in \Gott.\footnote{ \BT's November 24th, 1946 letter to his parents.}
 As it turned out,  more time was needed before the offer could be approved by the University administration, and in late September he was back in \Gott, where
 the landlady refused to let him back to his room, and he had to sleep on hard pavement, 
 until, presumably, rescued by his friends.  

During the summer and the months to follow, 
Bruno  worked hard on double beta decay.\footnote{November 24th, 1946, letter to parents from \Gott.}   In those months,   traveling between different places and countries did not permit  easy  concentration on  physics. Still,  he worked on the problem  while in the UK and started writing a paper, which was then submitted for publication.  Upon his return to \Gott, focusing better on his physics,  he found there was a mistake in his conclusions, and had to chase the error   to  correct it  before the paper would be published \citep{Touschek:1948ab}.\footnote{The paper, submitted  to {\it Zeitschrift f\"ur Physik} (now  the {\it European Physical Journal}), on December 2nd, 1946,  was published  in 1948. In this article  Touschek  thanks Heisenberg for suggesting the problem and for advice.} 
The anxiety about correcting an error,  trace  its origin, and rushing to have it corrected while the article was under publication, 
took most of his energies in October and November. 
In addition he had to move away from the betatron affairs, where  some  British officers were still keen on obtaining work or informations from him. 

\begin{figure}[htb]
\centering
\includegraphics[scale=0.25]{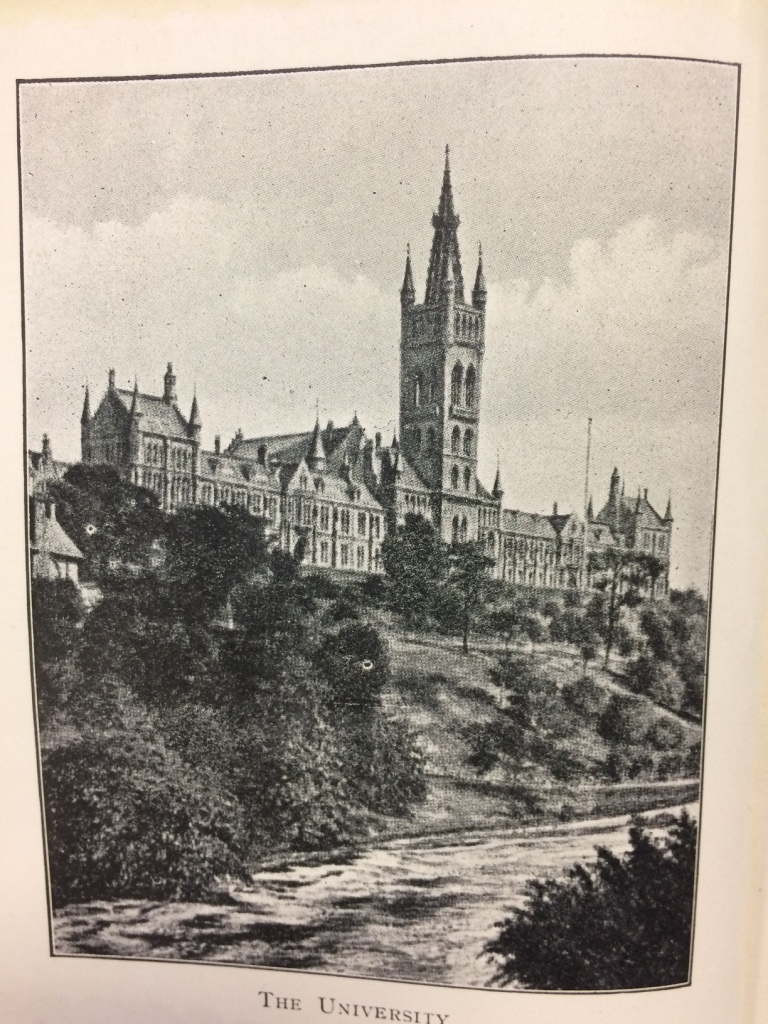}
 \hspace{1cm}
\includegraphics[scale=0.25]{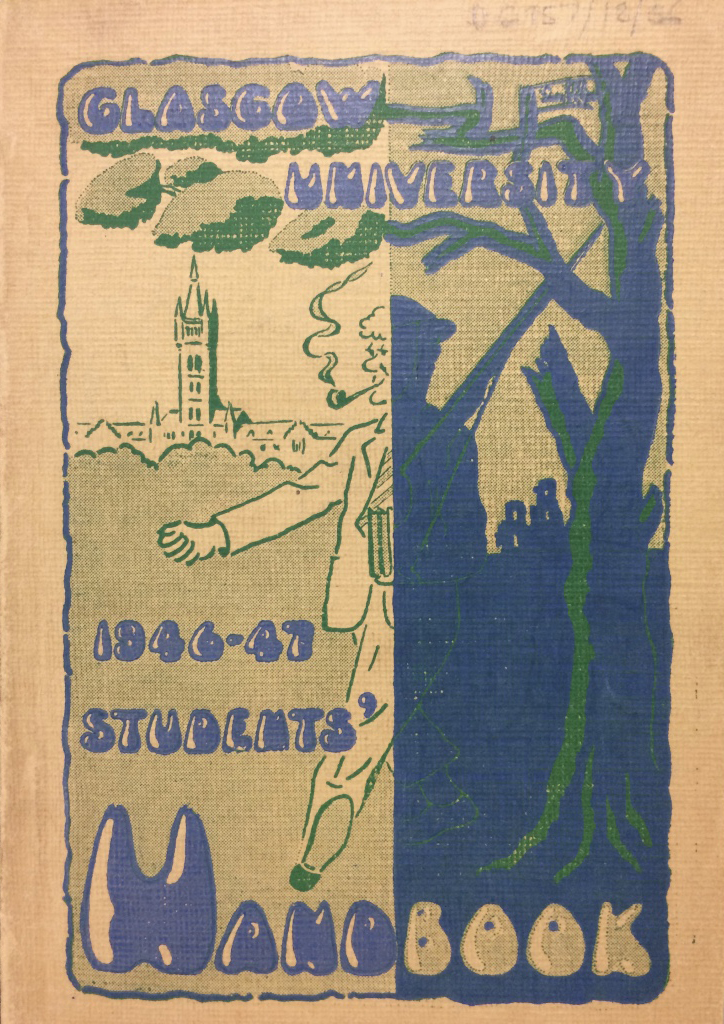}

\caption{
Left and right panels reproduce  two pages from the University of Glasgow 1946-47 Student Handbook, an image of the University and the front page, photographed from 
the original booklet, University of Glasgow Archives \& Special Collections, University collection, GB 248 DC 157/18/56. Reproduced with permission from University of Glasgow Archives.}
\label{fig:studetHandbook}
\end{figure}

Not receiving news from Glasgow, the uncertainty about where he would be in the next year became a pressing concern and,  on November 5th, Touschek solicited Philip Dee for an answer about the Glasgow position.\footnote{November 14th, 1946, Philip Dee's letter to \BT, Bruno Touschek Archive, Box 1.} During  these months,   Philip Dee, who was  keen  on having  Touschek  come to Glasgow,  had  continued  his efforts on Touschek's behalf, for him to  come to Glasgow, %
 and  enter the University Doctoral program.  In Fig.~\ref{fig:studetHandbook} we show two images from 
 the 1946-47 Student Handbook of the University of Glasgow.
 
The solution was near, but it  would take another four months before Touschek could take on his 
 research fellowship in Glasgow.

Touschek's anxiety  about his  future was also entangled with a degree of uncertainty 
  about the direction his research should take. He saw that  purely theoretical  problems  were  not interesting him any more, and felt he perhaps lacked the enthusiasm to persevere and solve them. He went to Heisenberg for advice, but could still not see his way out.  Various other difficulties piled up, including financial ones. At the end  of November, after Dee's letter, the only strategy  for Bruno  appeared  to  let the British authorities take care of his next move, although it was clear to him that no  solution would  be the perfect one. 
 Behind  uncertainty and doubts, there looms large  the presence  of Werner Heisenberg, who befriended Touschek, and may have been his inner mentor throughout his life. 
 
 Heisenberg was one of the great scientists who constructed the theoretical framework sustaining 
particle physics, a concerned  observer of the influence of science and philosophy, and a controversial protagonist of the debate about the moral imperative of a scientist facing   political power. He  was also a  major influence on Touschek's development as a physicist.  Touschek and Heisenberg never collaborated on an actual paper, nor was Touschek to be his doctoral student.   However, they often discussed  physics together   and \BT \ occasionally worked on some problems of interest to Heisenberg:  when a scientist of Heisenberg's stature makes himself available to intellectual and physics discussions, as in Touschek's  case, the effect   
 will last forever.\footnote{See the extensive correspondence between Heisenberg and Touschek preserved in Bruno Touschek's papers in Rome and in Heisenberg's papers at the Archive of the Max Planck Society in Berlin.} 
 The influence of Werner Heisenberg on Touschek runs deep through  Bruno's  work  with Walter Thirring \citep{Thirring:1951cz} on the Bloch and Nordsieck theorem  \citep{Bloch:1937pw} and in  statistical mechanics \citep{Touschek:1970aa}.   No matter how short,  six months or one year, 
 the contact with genius, when the latter allows it,    touches  one's mind and heart.


By mid December,\footnote{December 18th, 1946, letter to parents from \Gott.}
   Dee clarified that 
  the `unfortunate delay' was that all those involved in the affair  had forgotten that difficulties could arise at the University level -- not just at the  D.S.I.R.
  Once this was understood,   it had  then been necessary to wait for   the rectorate decision. This having been favourable to Touschek's hiring, 
it was now mostly a question for the appointment to go through  the usual official channels. This  would  naturally take some time, but it was now only   a matter of  few months.   This delay would suit Bruno, who was keen on attending a lecture by Heisenberg, to be held in January.

Once the Glasgow position had a definite starting date, April 1st, 1947,   Touschek could  see a clear way ahead of him, and could make closure with some of his past. In particular, he had to put an end to his parents' pressure to go back to Vienna. He had to definitely let his parents know that he would not look for a position there, as they were rather naturally asking  of him. Going to Glasgow was a clean break from the past. The  lost time was  his to reclaim, he hoped: the   five  years spent in   semi-hiding in Germany,  the two years between the Anschluss and the expulsion from the University of Vienna in June 1940, studying at Urban's home with borrowed books in 1941, all that lost time could be  retrieved. He was going to begin a new  life, and could  not afford to make any more mistakes. He would not go back to Vienna, at least not until he had his Doctorate.

 He saw that the first mistake had  been not to leave  Austria in 1938 or 1939, when, from Rome,  he had applied for a {\it visa} to go England. Waiting for it to arrive any day, ultimately he had returned to Vienna. Did he receive the {\it visa} but   lacked  the courage to go, or,  perhaps,  was the family support  not forthcoming?  It is quite possible that the difficulty may have been on missing family support.   In Vienna, they were  still hopeful for the worst not to happen. But it  was not going to be, as we know.  Bruno's  grandmother Weltmann, who had moved to Rome to stay with her  daughter Ada in 1938,    had later returned  to Vienna, after  Vittorio Emanuele III,  King of Italy, had signed into law the anti-semitic  regulations  declared by   Mussolini's regime in October 1938. But once in Vienna, around 1941  she had been taken to Theresienstadt, and never came back \citep[13]{Amaldi:1981}.

 
 In December 1946, when  Bruno's  diploma thesis was in print and the Glasgow situation was clearly in sight,  Bruno could sever the bond with his native city. 
He saw that  offers for a position in Vienna were not forthcoming:  if he were to go back,  he would be one of the clamoring many, and would need to enter into the typical academic squabbling and competition, something he did not,  and  would never have, appetite for.
As for his next move, he had no doubt that, from the scientific and intellectual point of view,   remaining in \Gott \  would be the most favourable way to go ahead towards a doctorate, but this was not to be taken for granted, not to mention the poor financial prospects. 
In fact  the financial situation of Heisenberg's Institute was still a difficult one, with scarce possibilites to support 
PhD students.

The problem of money  
was  a natural consequence of Touschek's rather desperate economic situation for a number of years.  From a rather affluent pre-war, pre-Anschluss life, he had been thrown  into the need to support himself when  semi-hiding in Germany. The war over, one can see the emergence of 
a moral imperative to support his parents  in Vienna,  under Soviet occupation.
Touschek's father had been a major in the Austrian Army and was now  retired, and  Touschek believed that his father had  perhaps left the Army under pressure because of   him, his   Jewish son, from his first marriage. In later conversations, with Edoardo Amaldi and Carlo Bernardini, his closest friends during the twenty five years he lived in Italy, Touschek let transpire a feeling of guilt in this respect \citep{Amaldi:1981}.  None of this can obviously be found  in Touschek's writings, 
but the {\it leit motif} of financial concern, and how much he could help his father and his step mother  is omnipresent. 
Thus, in April 1947, Touschek joined the Physics Department of the second most ancient of Scottish Universities, as a doctoral student in Glasgow. 
In later years,  Touschek regretted not  having  remained in Germany, but the history of science tells us that this was the right decision. In Glasgow,  Touschek would develop into a full fledged theoretical physicist, and establish  contact with the young Italian theorist Bruno Ferretti,  who would bring him to Rome in December 1952, where one of the great adventures of particle physics were to begin {a few years later}. There, in the nearby hills overlooking the  city spreading down to the Thyrrenian Sea, a new laboratory would be conceived in 1954 and built, and an electron synchrotron constructed and made to operate in 1959.    In this laboratory, on February 17th, 1960,  Touschek  proposed to  construct AdA, an  electron-positron collider, the first storage ring of matter-antimatter particles   in a laboratory  \citep{Amaldi:1981,Bonolis:2011wa}. Through AdA's operation and first successes,  there  came   the development of a new type of accelerator, which,  in the fifty years to follow, would  unravel  many of the mysteries of the world of particle physics.

%
\section{Who made the decision for Touschek's move to Glasgow in 1947: T-Force, Touschek or Heisenberg himself?}
\label{sec:changeofplans}
The sudden change of plans in April 1946, when Touschek first went to Glasgow and, one week later, left and went back to G\"ottingen to get his diploma, 
can be understood if we place Touschek's personal story in the wider context of how the Allies were planning for the scientific and technological future of the Western world, in a race against the Russians. We have seen the development of Operation Epsilon, through which the German nuclear scientists were chased and brought to England,  to be kept without any contact with family and colleagues for six months.  In January 1946 they were released to 
return to Germany, where they would  rebuild  German science in its  less war related aspects, namely no applied nuclear physics, no new accelerators, and other restrictions. Much more sinister, and better    known in its general lines, was  Operation Paperclip, which brought to the United States many scientists involved in rocket building, in chemical and biological warfare as detailed in \citep{Jacobsen:2014}.\footnote{Chief among the German scientists brought to the US was Wernher von Braun. Main scientist of the Nazi rocket program, including  the V-2, he became  the main artifex of the American space program,   as director of NASA's Marshall Space Flight Center and chief architect of the Saturn V launch vehicle which propelled the US to the Moon (see biography of Wernher von Braun at \url{https://history.msfc.nasa.gov/vonbraun/bio.html}). Bringing a number of German scientists, some of whom turned out to have been directly involved with slave labor in the concentration camps during the war, often gave rise to contrasts between the military and the US Immigration and Naturalization Service, most of the times resolved in favor of the military by higher political decisions.}

In the context of our story, we should recall that  the October 1945  B.I.O.S. report about Wider\o e's betatron had recommended that Bruno Touschek be brought to the UK.\footnote{B.I.O.S. Miscellaneous Report No. 77, Technical Report No. 331-45, European Electron Induction Accelerators.}
 This was also what he mostly wished to happen at the time. As 1946 rolled in, we have also seen that in January  a program for constructing  new particle accelerators was proposed by the  UK
Nuclear Physics subcommittee of the Government Advisory Committee on Atomic Energy. This program  was then {endorsed by the Committee on March 28th, 1946, and, shortly after, approved by the UK government \citep[491]{Krige:1989aa}. } 
This is why Ronald Fraser was able to carry through Touschek's proposed hiring in Glasgow, where he would  complete  his studies and eventually get his doctorate. No time seems to have been  wasted after the UK Government approved the construction of the new accelerators, and  in April  Touschek was brought  in the UK by the military to start his work in Glasgow.  Apparently, 
the immigration authorities, at Harwich, had some objections and officially refused landing rights.  
However,   
this did not stop Touschek  from entering the UK, something which  had also happened before, during a first visit in January or February, but  the military was able to override the civil authorities. However, when the  Darwin fellowship could not be approved because of him being Austrian rather than German,  nothing could be done, and he went back to G\"ottingen, to continue his studies there.

On June 26th Touschek obtained his diploma, which had been  a great success as   Sommerfeld wrote in a letter to Paul   Urban.\footnote{Postcard from Sommerfeld to Urban, in Amaldi Archive, Sapienza University of Rome, Box 524, Folder 4, Subfolder 4. It was probably a document sent by Urban to Amaldi, when the latter was preparing his biography of Touschek \citep{Amaldi:1981}.} We can now see various  parallel actions being set in motion. While  Dee and Fraser were trying to get him to join Glasgow, Touschek, emboldened by  his diploma,
was  now hopeful to remain in G\"ottigen and do his doctorate with Heisenberg. Other options were also open, but the most coveted would obviously be to remain and work with Heisenberg. It did not happen. He did receive a six month position, but that was all. 
We  have no  hints regarding Heisenberg's intentions about keeping Touschek at his institute, but  at the end of 1946/early 1947, the Kaiser Wilhelm Institute for Physics 
was still in a very difficult phase, most probably not yet in the position of funding PhD students. 
Touschek discussed his prospects  at length with Fraser, who would assure him things would be OK if he, Bruno, would remain in Germany. Was this all as straightforward as it appears? Or did the T-Force decision that Bruno was needed in Glasgow influence Heisenberg so that Touschek's only way forward was to go  to Scotland? Did Bruno ever {have}   a different choice?  We may  never know,  but the background story is so much larger than what Touschek could  see, that various possibilities co-exist. Once the accelerator program was approved by the UK government in March 1946, his move to Glasgow had to take place one way or the other. Dee (and Fraser)
could not immediately overcome the  obstacles posed by the civil authorities, but eventually they did, and Touschek (and his professors in G\"ottingen) had no choice. Namely, from the very beginning, it is very likely  Touschek was  meant to   take  the Glasgow   way,   because his expertise was of interest  to  the British scientists planning for the future of particle physics in the UK.

It may appear that we 
are assigning too much importance to Touschek  in this context, but one cannot forget the exceptional intellectual qualities that he possessed and were clearly seen by his peers, Arnold Sommerfeld, or Max Born, among them: coupled with  the unique experience with Wider\o e,  this combination is what ultimately led to the success of AdA. In Touschek, one finds the  potential for  innovation and disruption: he was  a theoretical physicist who had learnt   the ways of electrons, during the dark days of World War II, under the guidance of \RW,   the European authority  on  electron accelerators at the time. Thanks to such combination, of theory and practical expertise, in due time, \BT \    could envisage and  build a new type of accelerator, a matter-antimatter collider.

\section*{Acknowledgements}
G.P. gratefully acknowledges hospitality at the MIT Center for Theoretical Physics, and useful conversations with Prof. E. Lomon. We thank  Sapienza University of Rome, Deutsches Museum and the Max Planck Society for access to their Archives and are grateful to  Glasgow University Archives for allowing reproduction of the images  from  the 1946-47 Student Handbook. We thank  Neelam Srivastava and Amrit Srivastava for  assistance in  copy editing of the manuscript, and  
acknowledge the collaboration of  Amrit Srivastava for translating some of the cited documents. We are grateful to Angelo Mainardi and Yogendra Srivastava for constant encouragement,  comments and suggested  corrections to the text. One of us   thanks  Dorit Strauss for helpful discussions and advice. 
  \bibliographystyle{abbrvnat}
\bibliography{Touschek_book_Oct_20}
\end{document}